# Photon-Counting Interferometry to Detect Geontropic Space-Time Fluctuations with GQuEST

Sander M. Vermeulen[1,*] Torrey Cullen[1] Daniel Grass[1] Ian A. O. MacMillan[1] Alexander J. Ramirez[1] Jeffrey Wack[1] Boris Korzh[2] Vincent S. H. Lee[1] Kathryn M. Zurek[1] Chris Stoughton[3] and Lee McCuller[1]

[1]*Division of Physics, Mathematics and Astronomy, California Institute of Technology, Pasadena, California 91125, USA*
[2]*NASA Jet Propulsion Laboratory, Pasadena, California 91109, USA*
[3]*Fermi National Accelerator Laboratory, Batavia, Illinois 60510, USA*

The gravity from the quantum entanglement of space-time (GQuEST) experiment uses tabletop-scale Michelson laser interferometers to probe for fluctuations in space-time. We present a practicable interferometer design featuring a novel photon-counting readout method that provides unprecedented sensitivity, as it is not subject to the interferometric standard quantum limit. We evaluate the potential of this design to measure space-time fluctuations motivated by recent "geontropic" quantum gravity models. The accelerated accrual of Fisher information offered by the photon-counting readout enables GQuEST to detect the predicted quantum gravity phenomena within measurement times at least 100 times shorter than equivalent conventional interferometers. The GQuEST design, thus, enables a fast and sensitive search for signatures of quantum gravity in a laboratory-scale experiment.

## I. INTRODUCTION

The aim of the study of quantum gravity is to find a description of gravitation in concordance with quantum mechanics. Quantum gravity research is challenged by the vast difference in the scale of theoretically predicted gravity phenomena and the scale of quantum phenomena that can be measured in experiments. However, a number of different theories propose that the quantum nature of gravity implies fluctuations of the space-time metric that accumulate over macroscopic distances and yield potentially measurable uncertainties [1–5].

We focus here on "geontropic" fluctuations as proposed by Verlinde and Zurek [4], so named because they are fluctuations of space-time geometry induced by entropy. This theory provides a concrete prediction for a quantum gravity signal expected in interferometers. These fluctuations of space-time geometry are associated with quantum degrees of freedom counted by entanglement entropy, and the magnitude of the measured length fluctuations scales with the macroscopic size of the experiment (i.e., the "IR" length scale).

When a photon propagates in a metric that exhibits these fluctuations, it accumulates a change of phase compared to the case without fluctuations. When one compares the phase of two photons propagating along different paths in this randomly fluctuating metric, as in a Michelson interferometer, the observed phase difference varies in a way characteristic of the space-time fluctuations.

In Sec. II A, we review the theoretical motivation for these fluctuations in the metric from quantum gravity, and in Sec. II B we review how a signal from geontropic space-time fluctuations appears in interferometers. In Sec. III, we present our experimental approach of using a Michelson interferometer with a novel "photon-counting" readout method to detect this signal and compare the Fisher information [or the "signal-to-noise ratio"; see Eq. (6)] provided by this new approach with that of the standard "homodyne readout" scheme. We present our photon-counting interferometer design and projected sensitivity in Sec. IV. Section V discusses our strategy of a staged construction of the experiment, which is designed to positively identify a quantum gravity signal and distinguish it from known effects. We conclude in Sec. VII. A detailed consideration of the experimental challenges and sources of noise is included in Appendix A.

## II. THEORETICAL MOTIVATION

### A. Quantum gravity model

The status of theoretical results building on the work of Verlinde and Zurek *et al.* (VZ) up to the year 2022 is summarized in Ref. [6], showing that diverse theoretical approaches predict metric fluctuations of the same scale. The pixellon model was proposed in Ref. [7] to give detailed predictions for interferometric measurements based on the general theoretical expectations proposed by VZ. References [8,9] describe how shock-wave geometries give rise to these fluctuations. Details of the sensitivity of interferometers to geontropic fluctuations modeled by the pixellon are described in Ref. [10] with several testable predictions for the power spectral density, angular correlations, and low-frequency (IR) cutoff of the signal. It was also found in Ref. [11] that geontropic fluctuations would severely impact the sensitivity of future gravitational-wave detectors.

[*]Contact author: smv@caltech.edu

The quantum gravity theory behind VZ geontropic fluctuations incorporates the entanglement of quantum states on surfaces that define regions of space (see, e.g., Refs. [12–14]). These states are not directly observable. However, an essential conclusion of the theory is accessible to experimental tests: the presence of an isotropic, spherical breathing perturbation of the metric, which can be described by the pixellon scalar field $\phi$ [7,10]:

$$ds^2 = -dt^2 + (1 - \phi)(dr^2 + r^2 d\Omega^2). \quad (1)$$

The theory by VZ proposes degrees of freedom that fluctuate from entanglement entropy, and this scalar field represents the influence of those degrees on gravitation. This field $\phi$ is predicted to obey a wave equation and have a thermal distribution with Bose-Einstein statistics [7]. The gravity from the quantum entanglement of space-time (GQuEST) experiment will measure or constrain the metric fluctuations from the scalar field $\phi$.

Theories of geontropic fluctuations consistently determine that the scale of rms length fluctuations accumulated over a distance $L$ is given by

$$\langle \delta L^2 \rangle = \alpha \frac{l_p L}{4\pi} \approx \alpha (5.7 \times 10^{-18} \text{ m})^2 \left(\frac{L}{5 \text{ m}}\right), \quad (2)$$

where we use a convention for the Planck length of $l_p = \sqrt{8\pi\hbar G/c^3} = 8.1 \times 10^{-35}$ m. For reference, we normalize $L$ to an experimental scale of 5 m. Theoretical uncertainty in the fluctuation magnitude is encapsulated in the parameter $\alpha$ (note $\langle \phi^2 \rangle \propto \alpha$). Of particular note is that diverse approaches to quantum gravity yield $\alpha = \mathcal{O}(1)$. These include analyses from conformal field theory [14], dilaton theory [15], and hydrodynamics [16]. We expect that, with further development of the theoretical tools, $\alpha$ will be calculated exactly.

We take $y = \mathcal{O}(x)$ to denote an order-of-magnitude estimate of $y$, specifically, $y = \mathcal{O}(x) \Rightarrow 10^{-0.5}x < y < 10^{+0.5}x$. Our use of the approximation symbol $\approx$ indicates accuracy to the number of expressed significant digits.

### B. Interferometer signal from geontropic fluctuations

A laser interferometer (IFO) uses laser light to measure the accumulated phase difference between light traversing two arms. The phase accumulation due to geontropic fluctuations is stochastic and must be described statistically. We use the power spectral density (PSD) of the phase differences that geontropic fluctuations impart on the light to motivate, design, and benchmark experimental tests. We express the PSD $S_L^\phi(f)$ in terms of the equivalent optical path length differences that correspond to the measured phase fluctuations $\sqrt{\langle \delta L_{12}^2 \rangle} \equiv \sqrt{\langle (\delta L_1 - \delta L_2)^2 \rangle}$, where $\delta L_1$ and $\delta L_2$ are the effective optical path length changes of the individual arms; $\sqrt{\langle \delta L_{12}^2 \rangle}$ is not exactly equivalent to Eq. (2), as that expression gives the variance or autocorrelation for measuring a single length rather than the rms fluctuation of a differential length measurement of two nearby arms. This PSD $S_L^\phi(f)$ is the Fourier transform of the autocorrelation function of the differential length changes corresponding to accumulated phase differences of light returning to the beam splitter $\langle \delta L_{12}(t) \delta L_{12}(t-\tau) \rangle$, per the Wiener-Khinchin theorem. Here, $\tau$ is the time variable that is integrated over, and the autocorrelation function is invariant over all $t$. $S_L^\phi(f)$ is normalized as a single-sided PSD as a function of frequency $0 < f < \infty$.

The PSD is computed in Ref. [10] using the pixellon model [7]. This model is a low-energy effective description of the complete theory. The pixellon model PSD is depicted in Fig. 1. Note, however, that a *derivation* of the pixellon model from a full theory (i.e., one that also works at high energy, that is UV complete) is still underway. Several specific properties of this theoretical PSD are relevant for calculating experimental requirements. The peak level of the spectral density $\bar{S}_L^\phi \equiv S_L^\phi(f_{\text{pk}}) \geq S_L^\phi(f)$ is given by [10]

$$\bar{S}_L^\phi = \alpha \frac{l_p L^2}{8\pi^2 c} \approx \alpha \left(2.9 \times 10^{-22} \frac{\text{m}}{\sqrt{\text{Hz}}}\right)^2 \left(\frac{L}{5 \text{ m}}\right)^2. \quad (3)$$

The signal PSD scales with the measurement length as $L^2$, where one factor of $L$ arises from the signal bandwidth (see below). The PSD also scales with the free theoretical parameter $\alpha$, which we seek to measure or bound. Current theoretical expectation, based on Refs. [4,14,15,17], corresponds to $\alpha \lesssim 1$. Note that we denote the peak value or most representative level of the signal and other spectra with an overline ($\bar{S}$) from here on.

The peak of the spectrum is at $f_{\text{pk}}(L) = \mathcal{O}(c/[2\pi L])$, with a signal bandwidth $\Delta f(L) = \mathcal{O}(c/[2\pi L])$. From numerical evaluations of the PSD, we compute the specific values

$$f_{\text{pk}} \approx 15.6 \text{ MHz} \left(\frac{5 \text{ m}}{L}\right), \quad \Delta f \approx 36 \text{ MHz} \left(\frac{5 \text{ m}}{L}\right). \quad (4)$$

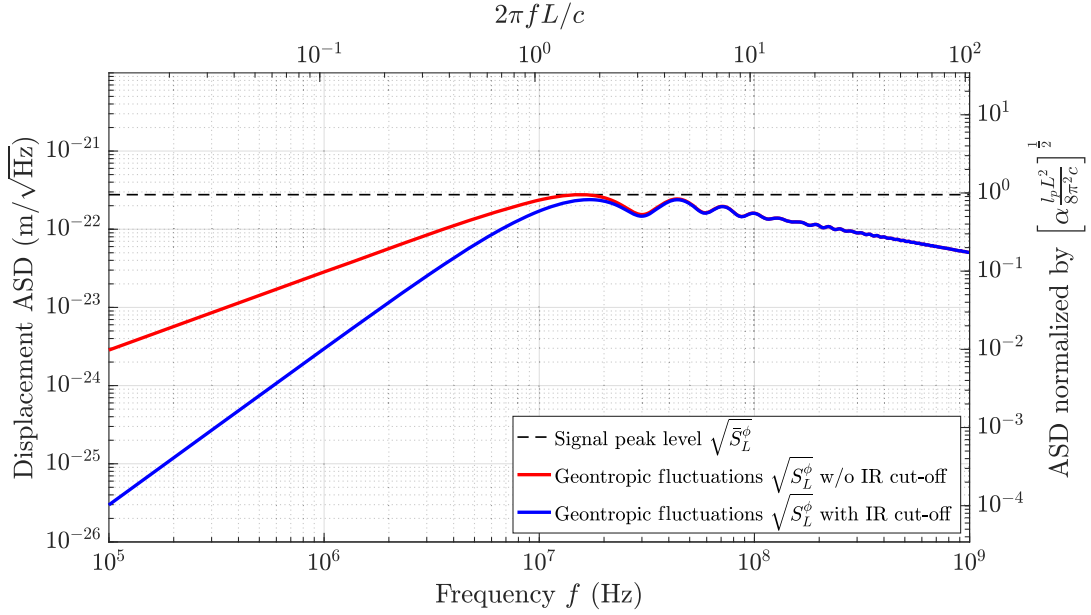

FIG. 1. The predicted displacement amplitude spectral density (ASD) signal due to geontropic fluctuations in the GQuEST experiment, assuming an arm length $L = 5$ m and $\alpha = 1$. The ASD is based on the pixellon model of geontropic fluctuations [10], and two different variations are shown, one where the signal spectrum has a low-frequency or IR cutoff at a frequency $c/L$ and one without this IR cutoff. The top $x$ axis shows the angular signal frequency $2\pi f$ normalized by the light-crossing frequency $c/L$. The $y$ axis on the right shows the signal ASD normalized to the fundamental scale $\alpha l_p L^2/(8\pi^2 c)$, as used in Ref. [10].

As is clear from Fig. 1, the signal is broadband with multiple peaks, and the definition of the signal bandwidth is somewhat arbitrary. The 3-dB FWHM bandwidth, which is approximately 16 MHz, is not suitable here; instead, we use $\Delta f \approx 36$ MHz as motivated from detection statistics in the following section.

We note that the signal amplitude depends on the angle between the two IFO arms $\Theta$. This is due to the spatial correlations of the geontropic fluctuations; fluctuations that affect the light in both arms coherently are canceled out, which suppresses the effective differential optical path length change signal measured at the IFO output. The signal amplitudes indicated in this work are for $\Theta = 90°$. The amplitude decreases to zero as $\Theta \to 0°$. The precise angular correlation is discussed in Refs. [4,10].

The use of two instruments to detect geontropic fluctuations can be advantageous, as the signal is expected to be correlated for collocated IFOs, and dominant noises are not. Two collocated IFOs have a frequency-dependent signal coherence $C^\phi(f)$ which depends on the separation between their beam splitters $L_{\text{sep}}$ and their relative orientation [11]. The coherence is defined as the ratio of the magnitude of the signal cross-spectral density to the geometric mean of the signal auto power spectral densities of the individual IFOs. Following the treatment in Ref. [11], we find that the coherence at $f_{\text{pk}}$ is $\bar{C}^\phi \equiv C^\phi(f_{\text{pk}}) = 0.88$ for $L_{\text{sep}}/L = 0.3$. This coherence factor is used again in Sec. IV H.

### C. Limits from existing experiments

Quantum space-time fluctuations have not yet been observed. However, existing interferometric experiments and astronomical observations impose tentative constraints on the phenomenology.

The LIGO interferometers are the most sensitive in terms of detectable strain fluctuations in their sensitive bandwidth [18]. However, these 4-km instruments have reduced sensitivity at their respective peak geontropic signal frequency [i.e., at $f_{\text{pk}}(4 \text{ km}) \approx 20$ kHz). The Fermilab Holometer comprised a pair of collocated 40-m IFOs and was built to be sensitive at frequencies on the order of $f_{\text{pk}}(40 \text{ m}) \approx 2$ MHz [19]. The strongest experimental constraints on the strength of the fluctuations $\alpha$, therefore, come from LIGO and Holometer measurements, which at $3\sigma$ significance are roughly $\alpha \lesssim 3$ and $\alpha \lesssim 0.7$ (with IR cutoff) and $\alpha \lesssim 0.1$ and $\alpha \lesssim 0.6$ (without IR cutoff), respectively [10].

An experiment similar to GQuEST, called QUEST [20], is currently being commissioned at Cardiff University. QUEST comprises a pair of collocated tabletop IFOs using homodyne readout and is designed to exceed the sensitivity of the Holometer by using higher optical powers and by using squeezed states of light. At its proposed sensitivity, QUEST could probe values of $\alpha < 0.6$ (with or without IR cutoff) with $3\sigma$ significance in roughly five months of observation time.

Space-time fluctuations could potentially manifest in experimental observations other than interferometric measurements. For instance, images of distant astronomical objects should appear blurred as the phase front of the light is distorted by space-time fluctuations. By analyzing data from astronomical observations, constraints were set on the strength of potential space-time fluctuations [21,22] subject to important caveats [4,23]. Importantly, the transverse correlations of geontropic fluctuations render these constraints inapplicable to the theory considered here [24].

### III. EXPERIMENTAL APPROACH

In this section, we expound the advantage of photon-counting interferometry; we first review the basics of laser interferometry (Sec. III A) and homodyne readout (Sec. III B), after which we quantify the reduction of measurement time for the detection of geontropic fluctuations provided by photon-counting readout (Sec. III C).

#### A. Laser interferometry

A laser Michelson IFO operates by shining laser light at a beam splitter, which splits light into two arms delimited by end mirrors; the end mirrors reflect the light back, which is then recombined at the beam splitter, where it interferes. Any signal that perturbs the optical path length of light traveling inside an interferometer causes a phase modulation of the light in the arm. This can equivalently be described as the conversion of input laser light to light with frequency components that are offset from the source frequency; the resulting frequency components of the optical field are typically called sidebands. For a signal perturbation (i.e., a modulation) at frequency $f$, the input laser field, also called the carrier field, at frequency $\nu = c/\lambda = ck$ (where $\lambda$ and $k$ are the laser wavelength and wave number, respectively) is modulated to create sideband fields at frequencies $\nu + f$ and $\nu - f$.

By introducing a static difference in the optical path lengths of the arms, the destructive interference at one port of the beam splitter becomes incomplete, which gives rise to "fringe" light at the IFO's output. Perturbations of the arm lengths then produce modulations of this light level, allowing the difference in arm length to be inferred by continuously monitoring the output light power $P_{\text{out}}$. Equivalently, the observed modulation of $P_{\text{out}}$ can be described as due to the beating of sideband fields with the fringe light field. This readout technique for interferometric signals is called "dc readout" or "fringe readout" in the interferometry community, as it uses the constant-intensity or dc (fringe) light as a local oscillator. The use of a local oscillator field makes this a form of optical homodyne readout, analogous to homodyne detection in radio and microwave electronic systems. We note that the term homodyne readout for Michelson interferometers often colloquially refers to the use of a separately supplied local oscillator to create a balanced homodyne readout [25]. Both dc fringe-offset light or a supplied local oscillator with balanced detectors implement homodyne readout of the quantum optical states emitted by the interferometer, and both methods are subject to the signal-to-noise ratio (SNR) analysis of this work. In the following, we assume only vacuum states are injected, not squeezing or more exotic quantum states. Further quantum enhancements are considered in Sec. VI.

The following sections calculate the quantum limits to resolving geontropic signals with an interferometer, first by using the dc (or fringe) readout method, henceforth referred to more generally as homodyne readout, and second by directly detecting the power in optical sidebands.

#### B. Homodyne readout

The local oscillator field light power $P_{\text{out}}$ randomly varies due to photon shot noise [26], limiting the ability to resolve small modulations of $P_{\text{out}}$ due to signals. When operating the IFO at near-perfect destructive interference, the shot-noise level, expressed as a one-sided spectral density of the equivalent differential arm length perturbations, does not depend on the choice of local oscillator power $P_{\text{out}}$, only on the circulating power on the beam splitter, $P_{\text{BS}}$. The standard quantum limit from shot noise (SQL) is [27,28]

$$\bar{S}_L^q = \frac{\hbar c}{2k P_{\text{BS}}} \approx \left(6.2 \times 10^{-19} \frac{\text{m}}{\sqrt{\text{Hz}}}\right)^2 \left(\frac{10\,\text{kW}}{P_{\text{BS}}}\right)\left(\frac{\lambda}{1550\,\text{nm}}\right). \tag{5}$$

The interferometric SQL can equivalently be described as arising due to the interaction of the circulating light field with the vacuum (see below). It applies to any form of homodyne readout of an IFO. We express Eq. (5) using the optical wavelength and power at the beam splitter from the reference design in Sec. IV.

This noise level $\bar{S}_L^q$ is only one of many noise contributions; the other noises arise from classical processes that create fluctuations of the IFO arm length or the phase and amplitude of the light. At the signal peak frequency, these classical noises can be engineered to be substantially smaller than the quantum noise $\bar{S}_L^q$. However, the classical noises are not negligible when photon counting, described in Sec. III C, is used.

The peak signal level $\bar{S}_L^\phi$ given by Eq. (3) is below the shot noise $\bar{S}_L^q$ given by Eq. (5) by 7 orders of magnitude. However, given sufficient measurement time, conventional interferometers using homodyne readout could eventually detect the signal. The geontropic fluctuations $\bar{S}_L^\phi$ manifest as a stochastic noiselike broadband displacement signal, and this can be detected as excess noise on top of the known quantum shot noise.

In this paper, we define the SNR as the ratio of the expectation value of an estimator $\hat{\alpha}_M$ of the signal strength $\alpha$, to the root-mean-square error $\sigma[\hat{\alpha}_M]$ of that estimator.

The SNR defined as such is, thus, a statistical quantity and not the ratio of two physical quantities *per se* [29]. The estimator $\hat{\alpha}_M$ is a function of the acquired data under a certain measurement scheme $M$. The expectation value of the estimator satisfies $\mathbb{E}[\hat{\alpha}_M] \equiv \alpha$ and is assumed to saturate the classical Cramér-Rao bound given the quantum measurement (i.e., the positive operator-valued measure) defined by $M$ [30]. Thus, the SNR is here formally defined as

$$\mathrm{SNR}_M \equiv \frac{\mathbb{E}[\hat{\alpha}_M]}{\sigma[\hat{\alpha}_M]} = \alpha\sqrt{\mathcal{I}_M[\alpha]}, \qquad (6)$$

where $\mathcal{I}_M[\alpha]$ is the total Fisher information accumulated on the parameter $\alpha$ in the data obtained using measurement scheme $M$. The second equality above implies the estimator saturates the classical Cramér-Rao bound. In the following, the square of the SNR is used as the figure of merit, as it is linear in the observing time. The SNR expressions we give below are more formally validated and related to the quantum Fisher information [30] in Appendix B.

The $\mathrm{SNR}^2$ in a search for excess noise due to geontropic fluctuations in a shot-noise-limited interferometer using homodyne readout is given by [28,31–33]

$$\begin{aligned}\mathrm{SNR}^2_{\mathrm{homodyne}} &= \int_0^T \int_0^\infty \left(\frac{S_L^\phi(f)}{\bar{S}_L^q(f)}\right)^2 \mathrm{d}f\mathrm{d}t \approx T\Delta f \left(\frac{\bar{S}_L^\phi}{\bar{S}_L^q}\right)^2 \\ &\approx \alpha^2 \left(\frac{T}{6\times 10^5~\mathrm{s}}\right)\left(\frac{P_{\mathrm{BS}}}{10~\mathrm{kW}}\right)^2\left(\frac{L}{5~\mathrm{m}}\right)^3,\end{aligned} \qquad (7)$$

where $T$ is the total measurement time and the time required to achieve a $\mathrm{SNR}^2 = 1$ or $1\sigma$ significance test for $\alpha = 1$ would be around 160 hours of continuous operation. A $\mathrm{SNR}^2 = 9$ or $3\sigma$ significance test for $\alpha = 1$ would then be around two months of continuous operation. Note that our definition of $\Delta f$ as stated above is chosen to make the approximation of Eq. (7) exact, to account for the specific spectral shape of the signal.

This suggests that a 5-m IFO using homodyne readout is a feasible means to search for this signal but would require significant measurement time. Additionally, confirming the presence of excess noise due to the diminutive geontropic fluctuations using a single interferometer with homodyne readout requires precise and stable calibrations of the shot-noise level, which are difficult to achieve.

### C. Photon counting

GQuEST will use the recently proposed technique of single-photon signal sideband readout [28], also called photon counting, to bypass the quantum shot-noise limit (i.e., the SQL) and achieve unprecedented sensitivities within relatively short measurement times [34]. The photon-counting method works by filtering the output light of the interferometer such that single photons carrying the signal of interest can be detected. This detection scheme outperforms homodyne readout, which is quantum-shot-noise-limited, in the detection of stochastic signals [30].

To explain the advantage of photon counting as proposed for the GQuEST experiment, we start by considering operating an IFO at perfect destructive interference. In this case, there is no local oscillator light at the output port of the IFO; any light observed at the output implies either the presence of a signal or the presence of some fluctuation that perturbs the interferometer arms. If the quantum gravity signal is weak and the classical noise is negligible, one may count single signal photons exiting the output port.

The geontropic length fluctuations produce effective differential interferometer arm length fluctuations $\langle \delta L_{12}^2 \rangle$. A change in the differential arm length produces a proportional change in the flux of photons $\dot{N}$ at the output port, with a constant of proportionality [28]

$$G \equiv \frac{\partial \dot{N}}{\partial \langle \delta L_{12}^2 \rangle} = \frac{kP_{\mathrm{BS}}}{\hbar c}, \qquad (8)$$

called the optical gain of the interferometer. The differential arm length changes due to geontropic fluctuations, thus, produce a signal photon flux $\dot{N}^\phi$ at the output port, which is given by

$$\dot{N}^\phi = G\langle\delta L_{12}^2\rangle = \frac{kP_{\mathrm{BS}}}{\hbar c}\langle\delta L_{12}^2\rangle. \qquad (9)$$

This total signal photon flux (photons $\cdot$ s$^{-1}$ = Hz) cannot yet be evaluated unequivocally using, e.g., Eq. (2), as the PSD of the pixellon $\phi$ signal falls off as $1/f$ (see Fig. 1), and, therefore, its integral $[\langle \delta L_{12}^2\rangle \equiv \int S_L^\phi(f)\mathrm{d}f]$ diverges logarithmically. This can be attributed to the lack of a high-frequency (UV) cutoff in the pixellon theory which further theoretical development should resolve. We evaluate the total photon flux of the signal within some finite detection bandwidth by integrating over the photon flux spectral density; this flux spectral density is

$$\mathcal{S}_{\dot N}^\phi(\epsilon) = G\frac{S_L^\phi(f)}{2} = \frac{S_L^\phi}{4\bar{S}_L^q} \quad \text{for } \epsilon = \pm f; \qquad (10)$$

this quantity represents the frequency decomposition of the signal sideband photon flux as a *two-sided* spectral density. We use the two-sided spectral density to evaluate the photon flux, as geontropic signals of frequency $f$ produce signal sidebands at optical frequencies $\nu - f$ and $\nu + f$ which can be separately measured. For this reason, we specifically use $\epsilon$ to denote measurements at an optical frequency shift $-\nu < \epsilon < \infty$, to distinguish it from measurements at signal frequency $0 < f < \infty$. The last equality in Eq. (10) is obtained from relating the optical gain to the shot-noise level as $\bar{S}_L^q = \frac{1}{2} \cdot G^{-1}$, which expresses that the vacuum state of the electromagnetic field (with an expectation value of $\frac{1}{2}$ quanta) produces spurious displacement signals $\bar{S}_L^q$ in the output of the interferometer.

To show the effectiveness of photon counting, we evaluate the signal photon flux due to geontropic fluctuations in a range of frequencies $\Delta f$ above and below the laser source frequency:

$$\dot{N}_{\text{peak}}^\phi = \int_{-\Delta f}^{\Delta f} \mathcal{S}_{\dot{N}}^\phi(\epsilon) d\epsilon = \alpha \cdot \mathcal{O}(1) \text{ Hz} \left(\frac{P_{\text{BS}}}{10 \text{ kW}}\right)\left(\frac{L}{5 \text{ m}}\right). \quad (11)$$

For a measurement where photons are counted over an interval $dt$, the number of accumulated signal photons is $dN = \dot{N} dt$. The variance of the number of accumulated photons $\sigma_{dN}^2$ in this time interval is determined by Poisson statistics; therefore, $\sigma_{dN}^2 = dN$. Thus, when counting signal photons in an IFO operated at perfect destructive interference without any classical noise, the SNR accumulates over time as

$$\text{SNR}_{\text{counts}}^2 = \int_0^T \frac{(dN_{\text{peak}}^\phi)^2}{dN_{\text{peak}}^\phi} = \int_0^T \dot{N}_{\text{peak}}^\phi dt$$
$$\approx T \Delta f \frac{\bar{S}_L^\phi}{2\bar{S}_L^q} \approx \alpha \left(\frac{T}{0.25 \text{ s}}\right)\left(\frac{P_{\text{BS}}}{10 \text{ kW}}\right)\left(\frac{L}{5 \text{ m}}\right), \quad (12)$$

where we approximate the spectrum as a constant equal to the peak value over the bandwidth of the signal $\Delta f$. For this approximation, we use our definition of $\Delta f$ as for the homodyne readout scheme above rather than making the approximation exact by redefining the bandwidth [which would be necessary to account for the different powers of $S_L^\phi$ in the integrands of Eqs. (7) and (12)].

Comparing Eqs. (7) and (12) indicates that reading out the interferometer by counting individual signal-carrying photons is fundamentally and profoundly more efficient than the usual homodyne readout [34]. Under ideal conditions, it requires less than a second to detect geontropic fluctuations with $\alpha = 1$ at $1\sigma$ significance, and even a $5\sigma$ test of the theory would take less than a minute.

In practice, this sensitivity cannot be achieved with current technology, as a realistic interferometer cannot be operated at perfect destructive interference for many reasons. There will always be small amounts of light at the output port of the interferometer due to imperfections in the optics and low-frequency length perturbations of the arms. These small amounts of light, also known as contrast defects, constitute a photon flux many orders of magnitude greater than the signal in Eq. (9) and would obscure it.

However, the condition of having no local oscillator or contrast defect light at the output can be emulated by filtering the light at the output port, removing unwanted optical power. This exploits the fact that the frequencies of the output optical field carrying the signal (the signal sideband) are different from the frequencies of the optical field from both the input laser and much of the classical noise. GQuEST will use optical cavities to strongly filter the output light, letting through only photons with frequencies corresponding to the desired signal.

We can model the effect of the cavities as a bandpass filter function $F(\epsilon - \epsilon_r)$, where $\epsilon_r$ is the readout frequency, which is set by choosing the resonant frequency of the cavities to be at a detuning $\epsilon_r$ from the carrier, and $0 < F \leq 1$, where $F(0) \approx 1$, $F(-\epsilon_r) \ll 1$, and the passband bandwidth $\Delta\epsilon \approx 25$ kHz (cf. Appendix A 9). By choosing $\epsilon_r = +f_{\text{pk}}$, signal photons at sideband frequencies $(f_{\text{pk}} - \Delta\epsilon/2) < \epsilon < (f_{\text{pk}} + \Delta\epsilon/2)$ are transmitted through the optical filter cavities, and photons at different sideband frequencies are rejected. The filtered signal photon flux is then

$$\dot{N}_{\text{pass}}^\phi = \int_{-\nu}^{\infty} \mathcal{S}_{\dot{N}}^\phi(\epsilon) F(\epsilon - \epsilon_r) d\epsilon$$
$$= \int_{\epsilon_r - \Delta\epsilon/2}^{\epsilon_r + \Delta\epsilon/2} \mathcal{S}_{\dot{N}}^\phi(\epsilon) d\epsilon \approx \frac{\Delta\epsilon \bar{S}_L^\phi}{4\bar{S}_L^q}$$
$$\approx \alpha \cdot 1.4 \times 10^{-3} \text{ Hz} \left(\frac{P_{\text{BS}}}{10 \text{ kW}}\right)\left(\frac{L}{5 \text{ m}}\right)^2 \left(\frac{\Delta\epsilon}{25 \text{ kHz}}\right), \quad (13)$$

which is valid for the realistic case that the readout bandwidth $(\epsilon_r - \Delta\epsilon/2) < \epsilon < (\epsilon_r + \Delta\epsilon/2)$ is chosen near the peak of the signal and $\Delta\epsilon \ll \Delta f$. The output flux in Eq. (13) considers only photons due to the signal in Eq. (3); in Sec. IV, we expand on the design of the experiment, including filter cavities to enable photon-counting readout, and then evaluate the experimental sensitivity with the presence of noise in Sec. IV G.

## IV. EXPERIMENTAL DESIGN

In this section, we detail the experimental design and projected sensitivity of GQuEST. The interferometer diagram in Fig. 2 indicates the essential elements of the GQuEST IFO design. The fiducial design parameters for the GQuEST IFOs are summarized in Table I. The estimated noise separated into contributions from various sources is shown in Fig. 3.

### A. Interferometer design

The IFO arm length is chosen to be 5 m, which balances the increase of the signal strength for longer arms ($\bar{S}_L^\phi \propto L^2$) with technical constraints on photon counting that favor having the peak of the signal spectrum at higher frequencies [note $f_{\text{pk}} \propto 1/L$, Eq. (4)], as discussed in Sec. IV G. We set the IFO interarm angle $\Theta = 90°$ for simplicity, as the increase in the signal magnitude using larger angles is not substantial. We use a laser wavelength of $\lambda = 1550$ nm to enable the use of silicon optics; silicon has favorable properties for the suppression of noise, as argued below, but is opaque to light of shorter conventional wavelengths such as 1064 nm. The use of silicon also takes advantage of the continuing development of optics for this wavelength for future gravitational-wave detectors [35].

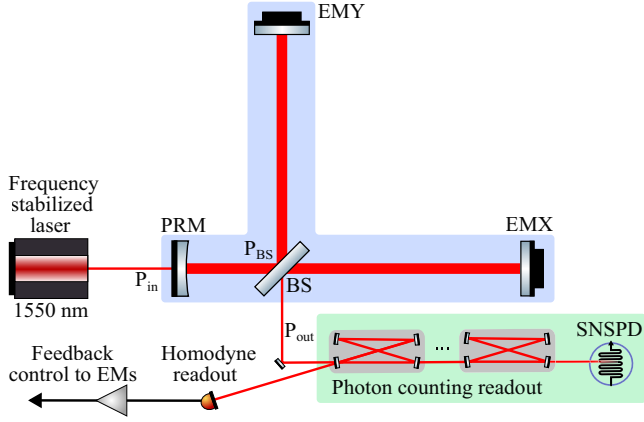

FIG. 2. Simplified diagram of the experimental design for one GQuEST IFO. Light from a frequency-stabilized laser with a wavelength of 1550 nm is incident on the beam splitter (BS) of the power-recycled IFO. The input laser light is transmitted through the power-recycling mirror (PRM), which together with the end mirrors (EMX and EMY) form the power-recycling cavity. The output light of the IFO is filtered through four narrow-band filter cavities, which provide > 200 dB of suppression of light at the laser frequency. Photons that pass through the filter cavities are detected using a superconducting nanowire single-photon detector (SNSPD). The light reflected off the first filter cavity is used in a homodyne readout scheme to enable feedback control of the IFO.

TABLE I. Parameters of the fiducial GQuEST design. The noise spectral densities are evaluated at 17.6 MHz.

| Parameter | Symbol | Value |
|---|---|---|
| Geontropic fluctuations scale parameter | $\alpha$ | $\mathcal{O}(1)$ |
| IFO arm length | $L$ | 5 m |
| Power on beam splitter | $P_{\mathrm{BS}}$ | 10 kW |
| Laser wavelength | $\lambda$ | 1550 nm |
| Laser frequency | $\nu$ | 193.4 THz |
| Nominal filter offset frequency | $\epsilon_{\mathrm{r}}$ | 17.6 MHz |
| Filter bandwidth | $\Delta\epsilon$ | 20 kHz |
| Twin IFO separation | $L_{\mathrm{sep}}$ | 1.5 m |
| IFO interarm angle | $\Theta$ | 90° |
| Signal spectral density (peak) | $\bar{S}_L^\phi$ | $\left(3 \times 10^{-22} \text{ m}/\sqrt{\text{Hz}}\right)^2$ |
| Thermal noise spectral density | $\bar{S}_L^c$ | $\left(10^{-21} \text{ m}/\sqrt{\text{Hz}}\right)^2$ |
| Shot-noise spectral density | $\bar{S}_L^q$ | $\left(6 \times 10^{-19} \text{ m}/\sqrt{\text{Hz}}\right)^2$ |
| Filtered signal photon flux | $\dot{N}_{\mathrm{pass}}^\phi$ | $1.4 \times 10^{-3}$ Hz |
| Filtered classical noise flux | $\dot{N}_{\mathrm{pass}}^c$ | $1.6 \times 10^{-2}$ Hz |
| Photon detector dark count rate | $\dot{N}^{\mathrm{d}}$ | $< 10^{-3}$ Hz |
| Observation time for $5\sigma$ test for $\alpha = 1$ | $T$ | $\mathcal{O}(10^5)$ s |

The interferometer is operated near destructive interference, allowing only a small fraction of the total power on the beam splitter to be directed toward the output port. The remainder returns toward the input laser. Between the input laser and the beam splitter, a power-recycling mirror is added, which forms a resonant cavity with the arm end mirrors (see Fig. 2). This power-recycling cavity enhances the injected laser power of 10 W to 10 kW or more of circulating light.

Low-frequency perturbations of the interferometer arms from the environment need to be counteracted to maintain the IFO at its operating point. This is done with feedback control, where the perturbations are read out by measuring modulations of the output field reflected off the first filter cavity (i.e., homodyne readout of the output power; see Fig. 2 and Appendix A 3).

The target output light power due to low-frequency perturbations and differential imperfections (i.e., the contrast defect) is $P_{\mathrm{out}} = \mathcal{O}(100)$ mW, which is small compared to the power on the beam splitter but large compared to the expected photon flux due to the geontropic signal. The following design elements enable the signal to be detected despite the presence of non-signal-carrying contrast defect light and other noise.

### B. Filter cavities

To suppress the contrast defect light, we use a series of narrow-band optical filter cavities at the interferometer output that resonantly transmit light at a frequency $\nu + \epsilon_{\mathrm{r}}$ (where $\nu$ is the frequency of the input laser and $\epsilon_{\mathrm{r}}$ is the signal sideband frequency). Based on the signal PSD in Sec. II A and the estimated noise PSD in Fig. 3 (see Sec. IV F), we choose a filter cavity offset frequency of $\epsilon_{\mathrm{r}} = 17.6$ MHz, with a filter FWHM bandwidth of 42 kHz (and a cavity pole of 21 kHz). Each of the four filter cavities, thus, provides roughly $20 \log(17.6 \text{ MHz}/21 \text{ kHz}) = 58$ dB of power suppression of the carrier light for a total of 232 dB of filtering, which reduces as much as 1 W of light at carrier frequency to a level below that of the signal. With multiple filters in series, the effective pass bandwidth is $\Delta\epsilon \approx 25$ kHz. During operation, the value of $\epsilon_{\mathrm{r}}$ can be varied almost arbitrarily, but sufficient filtering is expected to be achievable primarily in the range from 8 to 40 MHz (see Appendix A 9). Changing $\epsilon_{\mathrm{r}}$ allows the frequency dependence of the signal PSD to be resolved; this also enables the noise spectrum to be characterized.

### C. Single-photon detection with SNSPDs

Photons are detected downstream of the filter cavities using superconducting nanowire single-photon detectors (SNSPDs). SNSPDs have been demonstrated to achieve 98% detection efficiency at 1550 nm [36] and intrinsic dark count rates (i.e., the rate of spurious SNSPD signals in the absence of light) as low as $6 \times 10^{-6}$ counts per second (cps) [37].

These detectors are fabricated by patterning thin (approximately 5 nm) films of superconductors (typically, WSi, MoSi, NbN, or NbTiN) into nanowires in the region of 100–250 nm in width. This nanowire is meandered to cover the active area, where $20 \times 20$ μm is large enough to couple efficiently to a single optical mode at 1550 nm. To ensure high absorption in the device, the meandered nanowire is embedded into a dielectric stack, with either a metal [38] or dielectric [36] backreflector. The superconducting nanowire is current biased at a high fraction of its critical depairing current such that the absorption of the single 1550-nm photon is sufficient to break the superconductivity across the whole nanowire, through a highly nonlinear process [39,40]. The resistive domain in the nanowire rediverts the original current into a readout amplifier, providing a digital "click" to register the photon detection event, referred to as a "count."

While SNSPDs can have low intrinsic background count rates and high quantum efficiency using the techniques above, they must be optimized for use with an interferometer experiment. To maintain such low background count rates requires the output of the interferometer to be efficiently coupled to the SNSPD while preventing any spurious photons (e.g., from a thermal background) from producing counts in the detector. To reduce the thermal background, the optical fiber that carries the output photons to the SNSPD will be shrouded and cooled. In addition, a free-space coupling of the interferometer output to the SNSPD in a cryogenic environment might be required, although free-space coupled dark count rates of $10^{-2}$ cps have already been demonstrated for a $\lambda = 1550$ nm SNSPD readout [41].

### D. Thermal distortion from high optical powers

The principal advantage of the photon-counting readout is the elimination of quantum noise from the interferometric SQL [see Eq. (5)]. In the absence of this noise, the dominant noise encountered in the experiment is expected to be classical noise from thermal fluctuations in the optics.

An important means to mitigate this noise is the choice of optical substrate material. GQuEST will use crystalline silicon optics, instead of the fused silica used in other precision laser interferometers (e.g., LIGO and Holometer). Silicon has a higher thermal conductivity, a higher phonon propagation speed, and a higher mechanical quality factor than fused silica (at the operating temperature). These contribute to the reduction of different kinds of thermal noise (see Appendix A) [42]. Although silicon's absorption of light with a wavelength of 1550 nm is greater than the absorption of fused silica at 1064 nm, this effect is negligible, as the total absorption in the optics is dominated by absorption in the optical coatings.

In particular, the use of crystalline silicon mitigates thermal lensing in the beam splitter. Absorption of the light traversing the beam splitter creates a temperature gradient inside the substrate, which causes inhomogeneous refraction due to the temperature dependence of the index of refraction. This effect, known as thermal lensing, scatters light power from the fundamental Gaussian input mode into higher-order modes (see Appendix A 7) when the light is transmitted through the beam splitter. The resulting differential scattering between the arms perturbs the destructive interference at the output, creating a contrast defect. The contrast defect produces spurious output light at the carrier frequency that requires suppression by the filter cavities. If the fractional power leakage due to the contrast defect is large compared to the transmission of the power-recycling mirror, then it also limits the amount of power that can be built up in the interferometer.

### E. Laser noise

The output filter cavities remove the carrier light at frequency $\nu$, but the input laser light also carries noise spanning a range of frequencies. In addition to a 250 Hz laser linewidth, the input laser's spectrum has a white noise floor due to amplified spontaneous emission equivalent to a laser phase noise on the order of $10^{-7}$ rad/$\sqrt{\text{Hz}}$ [35]. This would amount to a photon flux of $\dot{N}_{\text{pass}}^{\text{LP}} = \mathcal{O}(10^8)$ Hz at the detector (after the readout filter cavities; see Appendix A 1), which is far greater than the signal photon flux and, therefore, requires suppression. An input filter cavity (which is elided and considered a part of the source in Fig. 2) and the power-recycling cavity each have a bandwidth $\mathcal{O}(10)$ kHz, and, therefore, each provides a power suppression of $\mathcal{O}(60$ dB) at the signal peak frequency for a total of $\mathcal{O}(120$ dB) of filtering of the laser amplitude and phase noise. The filtered laser phase noise spectrum is shown in Fig. 3. The suppressed laser noise photon flux is, thus, expected to be $\dot{N}_{\text{pass}}^{\text{LP}} = 10^{-4}$ Hz, which is negligible compared to the photon flux of the signal, calculated in Sec. IV G.

### F. Thermal noise

Sideband photons from the interferometer not due to the signal with frequencies in the filter passband ($\epsilon \approx \epsilon_r$) are detected as noise on the photodetector. The dominant source of such noise sidebands is expected to be thermal excitations of the optics, which couple to the circulating light in different ways.

Mechanical elastic resonances of the optical substrates entail oscillations of the reflecting surfaces of the optics. This produces significant noise sidebands at the frequency of the mechanical mode, with a frequency spread depending on the quality factor of the resonance. We model this mechanical noise analytically and numerically and find that, for thin, disklike mirrors, the mechanical modes (which we refer to as "solid normal modes") create a spectrum of noise peaks that are regularly spaced across the signal band, as plotted in Fig. 3 (see Appendix A 4) [19]. The spacing of the peaks, at our measurement frequencies, is determined by the speed of sound of (longitudinal) P waves in the material. The impact of these peaks is mitigated through the narrow bandwidth of the filter cavities: The filter passband is chosen to lie between successive mechanical resonances (see Fig. 3).

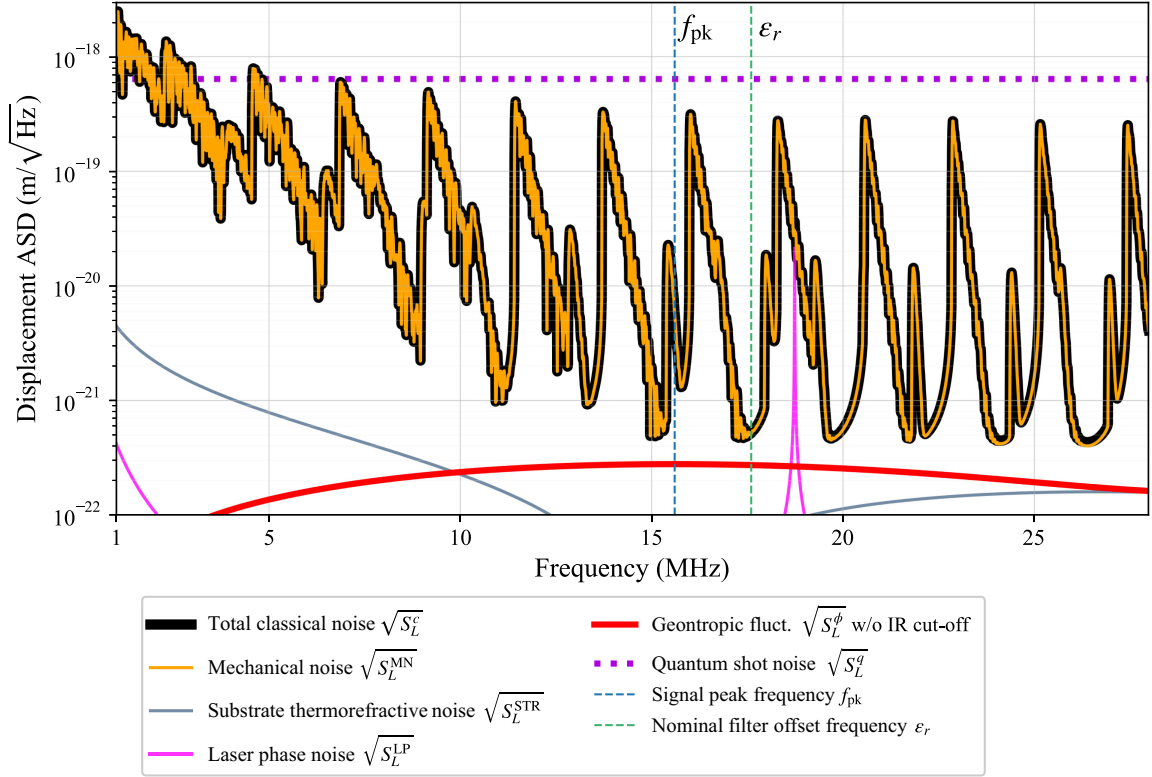

FIG. 3. The displacement amplitude spectral density of various dominant noises as estimated for the fiducial design (see Table I) are plotted together with the expected signal from geontropic quantum space-time fluctuations $\sqrt{S_L^\phi}$ (red line). The total classical noise $\sqrt{S_L^c}$, which limits the sensitivity of GQuEST, and its contributions are shown (orange, gray, and pink lines). The various noise contributions are considered in detail in Appendix A. The standard quantum limit from shot noise $\sqrt{S_L^q}$, which GQuEST is not subject to thanks to its photon-counting readout, is also shown (purple dotted line). The dark blue vertical dashed line marks the peak signal frequency $f_{pk}$. The green vertical dashed line marks the fiducial cavity offset frequency $\epsilon_r$; note that the filter bandwidth $\Delta\epsilon = 20$ kHz is narrower than the thickness of that line in this plot. See Fig. 5 in Appendix A for additional noise sources.

The optical coatings applied to the substrates entail additional thermal noise, and we model this effect (see Appendix A 4); the dominant contribution from coating thermal noise is expected to be its lowering of the total mechanical quality factor of the optics. Unlike previous modeling of coating thermal noise [43], which has implicitly considered only the frequency regime below the first mechanical eigenfrequency of the optics, this new model is applicable for frequencies in our readout band. Based on our model, it is expected that total mechanical noise (MN) (including substrate and coating) is the dominant contribution to the total classical noise, i.e., $S_L^{MN} \approx S_L^c = (5 \times 10^{-22} \text{ m}/\sqrt{\text{Hz}})^2$. For the reference sensitivity and measurement time estimates considered below, we therefore use a conservative reference level of $\bar{S}_L^c = (10^{-21} \text{ m}/\sqrt{\text{Hz}})^2 > S_L^c$.

We note that thermorefractive noise produced in the substrate of the beam splitter is expected to be a significant contribution to the total noise, and this noise source could even be dominant if a standing optical wave is formed in the beam splitter [44] as is the case in conventional interferometers. Therefore, GQuEST will use a modified optical configuration where no standing wave is formed to avoid this effect, as discussed in Appendix A 5.

Notably, the classical noise level is expected to be above the signal level, which implies a nonzero background photon count rate. Therefore, the sensitivity of the experiment will be limited both by the low flux of signal photons and by the variance of the flux of photons from thermal noise. The statistical impact and an experimental strategy to remove this noise are described in Secs. IV G and IV H, respectively.

### G. Reference sensitivity

To make a realistic estimate of the sensitivity of the interferometers, we have to evaluate the signal count rate

$\dot{N}^\phi_{\text{pass}}$, the count rate from classical interferometer noise, $\dot{N}^c_{\text{pass}}$, and the dark count rate of the photodetector, $\dot{N}^d$.

The filtered photon flux from classical noise is computed similarly to the computation of the filtered signal photon flux [see Eqs. (10) and (13)], where we substitute $S^\phi_L$ for the classical displacement spectral density $S^c_L$. In addition, for both the signal and the noise, we model the frequency dependence of the transmission of the optical filter cavities as the filter function $F(\epsilon - \epsilon_r)$. The filtered classical noise photon flux is then

$$\dot{N}^c_{\text{pass}} = \int_{-\infty}^{\infty} \mathcal{S}^c_{\dot{N}}(\epsilon) F(\epsilon - \epsilon_r) d\epsilon \approx \frac{\Delta\epsilon \bar{S}^c_L}{4\bar{S}^q_L}$$
$$\approx 1.6 \times 10^{-2} \text{ Hz} \left(\frac{P_{\text{BS}}}{10 \text{ kW}}\right) \left(\frac{\Delta\epsilon}{25 \text{ kHz}}\right), \quad (14)$$

which scales linearly with the expected classical noise level $\bar{S}^c_L = (10^{-21} \text{ m}/\sqrt{\text{Hz}})^2$ (see Appendix A 4) in the nominal readout bandwidth $\Delta\epsilon$ centered on the signal peak. The SNR can then be found by considering that the signal accumulates as $\int dt \dot{N}^\phi$, while the total variance is the quadrature sum of all noise count rate contributions, i.e., $\sigma^2_{dN} = \Sigma_i \sigma^2_{dN^i}$, integrated over time. This leads to an SNR of

$$\text{SNR}^2_{\text{counts}} = \int_0^T \frac{\left(\dot{N}^\phi_{\text{pass}} dt\right)^2}{\left(\dot{N}^\phi_{\text{pass}} + \dot{N}^c_{\text{pass}} + \dot{N}^d\right) dt}. \quad (15)$$

This can be evaluated as [28]

$$\text{SNR}^2_{\text{counts}} \approx \frac{T\Delta\epsilon}{4} \frac{\bar{S}^\phi_L}{\bar{S}^q_L} \left(1 + \frac{\bar{S}^c_L}{\bar{S}^\phi_L} + \frac{4\dot{N}^d}{\Delta\epsilon} \frac{\bar{S}^q_L}{\bar{S}^\phi_L}\right)^{-1}. \quad (16)$$

If the dark count rate and classical noise are negligible, the SNR is estimated to be

$$\text{SNR}^2_{\text{counts}} \approx \frac{T\Delta\epsilon}{4} \frac{\bar{S}^\phi_L}{\bar{S}^q_L} \quad (17)$$

$$\approx \alpha \left(\frac{T}{730 \text{ s}}\right) \left(\frac{P_{\text{BS}}}{10 \text{ kW}}\right) \left(\frac{L}{5 \text{ m}}\right)^2 \left(\frac{\Delta\epsilon}{25 \text{ kHz}}\right). \quad (18)$$

Here, the increase in the required measurement time compared to Eq. (12) is due to the reduced bandwidth of the readout filter cavities compared to the full signal bandwidth.

If we realistically incorporate that the classical noise is not negligible and is larger than the expected signal level ($\bar{S}^c_L > \bar{S}^\phi_L$) and additionally assume the dark count rate is negligible compared to the classical noise ($\dot{N}^d \ll \dot{N}^c$), the SNR is given by

$$\text{SNR}^2_{\text{counts}} \approx \frac{T\Delta\epsilon}{4} \frac{\left(\bar{S}^\phi_L\right)^2}{\bar{S}^q_L \bar{S}^c_L}$$
$$\approx \alpha^2 \left(\frac{T}{8500 \text{ s}}\right) \left(\frac{P_{\text{BS}}}{10 \text{ kW}}\right) \left(\frac{L}{5 \text{ m}}\right)^4 \left(\frac{\Delta\epsilon}{25 \text{ kHz}}\right). \quad (19)$$

From Eq. (19), we can infer that, at the design sensitivity, GQuEST will be able to probe values of $\alpha < 0.6$ at $3\sigma$ significance in 60 h of measurement time, which is the current experimental constraint set by the Holometer for geontropic fluctuations with IR cutoff. GQuEST can reach $\alpha < 0.1$ at $3\sigma$ in 2160 h, which allows it to go beyond the current LIGO constraint on the theory without IR cutoff.

The estimate of the detection statistic using a realistic photon-counting interferometer as given by Eq. (19) is a key result of this work. It should be compared against the detection statistic for homodyne (dc or fringe) readout as given by Eq. (7) and the detection statistic offered by an ideal photon-counting interferometer in Eq. (17). The rate of accrual of statistical power for a realistic photon-counting interferometer is proportional to $(4\bar{S}^q_L \bar{S}^c_L)^{-1}$, whereas the rate for homodyne readout is proportional to $(\bar{S}^q_L)^{-2}$. The factor of 4 appears in part from using only the positive component of the two signal sidebands and in part from the signal arising in only one of two optical quadratures. We note that the factor $\bar{S}^q_L$ that appears in Eq. (19) represents the optical gain $G = (2\bar{S}^q_L)^{-1}$, which establishes the rate that signal information is extracted as photons, while the factor $\bar{S}^c_L$ represents how classical background noise statistically slows the extraction of signal information.

Thus, three major ways to increase the sensitivity and decrease the required measurement time are to reduce the classical noise, increase the circulating laser power, or increase the arm length. The circulating laser power is set by engineering limits, as detailed in Appendixes A 2 and A 7. Increasing the arm length has the effect of shifting the peak signal frequency to lower frequencies [see Eq. (4)]. Importantly, at lower frequencies, the dominant classical noises will be stronger, as the thermal noise of the optics scales as $1/f$. Additionally, a subdominant noise source might become of influence at lower frequencies; thermorefractive noise has a $1/f^2$ frequency dependence. Moreover, the achievable carrier power isolation is smaller at lower frequencies, as this suppression scales as $1/f^8$ (for the fiducial design using four readout filter cavities). At sufficiently low frequencies, the mechanical noise of the mirrors is no longer concentrated at specific frequencies (see Fig. 3), which means the noise PSD no longer exhibits significant local minima that make for suitable readout frequencies. For these reasons, a design with 5-m arms is chosen to balance the signal magnitude and classical noise levels at the signal peak.

### H. Coherent signal detection in twin interferometers

The dominant classical noise produces a greater photon flux than the geontropic fluctuation signal. Therefore, measurement of the underlying signal requires the subtraction of the classical noise. Using a single interferometer, this subtraction requires accurate characterization of the noise floor. Specifically, the expected photon flux in the absence of signal must be quantified with an uncertainty smaller than the magnitude of the signal. Moreover, the total classical noise needs to be measured such that the observed noise is independent of the signal and yet is representative of the noise that would be observed if the signal were present.

For the quantum space-time fluctuations we consider, the signals measured in two collocated interferometers are highly coherent, while the dominant noise is incoherent. This opens the possibility of using two interferometers to separate the correlated signal from the uncorrelated noise. This avoids the aforementioned challenges involved in removing backgrounds from the detected signal power.

An established method for doing this is to cross-correlate two collocated interferometers that use homodyne readout. In this method, a product of the electronic photodetector signals of the two interferometers is taken to compute the cross-correlation, which represents a direct estimate of the coherent signal magnitude [45,46]. Cross-correlation, therefore, provides a great practical advantage and improves the required integration time of Eq. (7) by a factor of 2 from the use of two instruments [28]. However, the homodyne readout method is subject to the standard quantum limit from shot noise, and this cross-correlation method does not achieve the fast detection times that photon counting provides [Eq. (19)].

When using a photon-counting readout, the phase information of the optical field is lost after the detection of individual photons. Therefore, to exploit the coherence of the signals, the phase of the optical fields coming out of the two interferometers must be compared before the detection is made.

Coherent signal detection with twin GQuEST interferometers will be done using the setup shown in Fig. 4. The output light of two IFOs interferes on a beam splitter, BS-C, such that a coherent signal common to both input ports of the beam splitter interferes constructively toward one readout port and destructively in the other. Incoherent optical fluctuations from background noise in the two IFOs will, on average, be split equally between both BS-C outputs. Therefore, any counts detected downstream of the beam splitter output that contains only uncorrelated noise (the "null" channel) can be subtracted from the counts measured downstream of the other port (the "signal" channel). The expectation value of difference between the null and signal channels of the correlated photon flux provides a direct estimate of the geontropic signal.

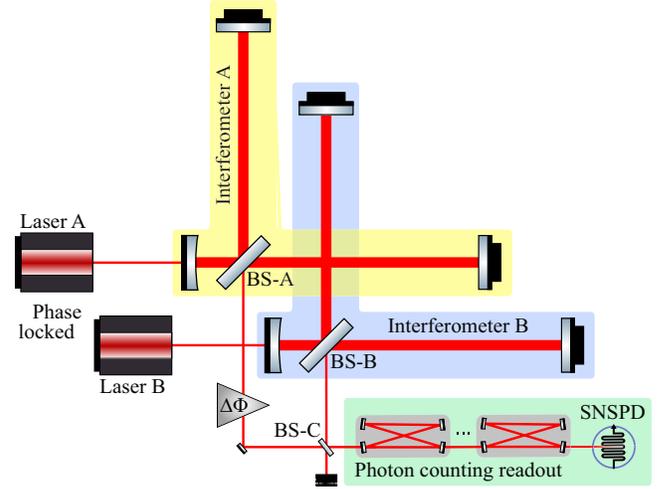

FIG. 4. Simplified schematic for the twin-interferometer (IFO) configuration of GQuEST. The outputs of two identical collocated IFOs are combined such that signals that are correlated between the IFOs interfere constructively in one output port of beam splitter BS-C and destructively in the other. Thus, a signal and a null readout channel can be created and swapped by adjusting the relative phase $\Delta\Phi$ of the two IFO outputs. To ensure the output signal is coherent, the two input lasers must be coherent (phase locked).

Which output port of BS-C contains the signal depends on the relative optical path lengths from the two interferometers to the beam splitter, and this can, therefore, be selected and switched as desired. This avoids the need to build two separate readout setups with filter cavities and SNSPDs downstream of the beam splitter, as the signal and null channels can be characterized separately at different times. The stationarity of the noise can be tested by switching between the signal and null channels over time.

In this coherent signal detection scheme, the SNR for excess photon counts in the signal channel is given by

$$\text{SNR}^2_{\text{coh counts}} = \int_0^T \frac{\left[\left(\dot{N}_{\text{signal}} - \dot{N}_{\text{null}}\right)dt\right]^2}{\left(\dot{N}_{\text{signal}} + \dot{N}_{\text{null}}\right)dt}$$

$$= \int_0^T \frac{\left(2\bar{C}^\phi \dot{N}^\phi_{\text{pass}} dt\right)^2}{2\left(\bar{C}^\phi \dot{N}^\phi_{\text{pass}} + \left(1-\bar{C}^\phi\right)\dot{N}^\phi_{\text{pass}} + \dot{N}^c_{\text{pass}} + \dot{N}^d\right)dt}, \quad (20)$$

where $\dot{N}_\text{signal}$ and $\dot{N}_\text{null}$ are the photon fluxes as measured using photon-counting readout in the signal and null channels, respectively. This expression assumes all excess photon counts in the signal channel are due to the coherent part of the signal. This SNR can be related to the SNR for a single photon-counting interferometer [Eq. (15)]:

$$\text{SNR}^2_\text{coh counts} = 2(\bar{C}^\phi)^2 \text{SNR}^2_\text{counts}. \quad (21)$$

This shows that, while coherent signal detection with twin IFOs provides a practical advantage as it does not require the characterization of the classical (uncorrelated) noise, it does not yield a major statistical advantage over a single photon-counting IFO. The statistic improves by a factor of $2(\bar{C}^\phi)^2$; this factor can be understood as quantifying the extra information provided by having two interferometers, which is reduced by imperfect coherence. However, when reading out only a single port of BS-C at a time, the factor of 2 improvement is lost.

We note that the number of excess counts in the signal channel could be biased due to a source of noise that is coherent between the ports of BS-C, due to correlations either upstream or downstream of BS-C. This noise would manifest as spurious signal power that needs to be characterized and subtracted away. As detailed in Appendix A 12, we anticipate that for the fiducial design the photon flux from all correlated noise sources are either smaller than the nominal signal photon flux $\dot{N}^\phi_\text{pass}(\alpha = \mathcal{O}(1))$, can be reduced to below this level, or can be characterized with an uncertainty smaller than this nominal signal photon flux.

## V. EXPERIMENTAL STAGES AND OPERATIONS

The final experimental stage of GQuEST will consist of taking data from two cross-correlated interferometers with the photon-counting readout described above. This ultimate experimental configuration of two high-power interferometers will require considerable development that can be performed in stages to demonstrate the requisite technologies, performance, and integration requirements needed to achieve our experimental goals. This is the preferred approach, as this experiment is likely to be limited by classical noises in frequency and magnitude regimes outside of those previously studied by the interferometry community.

### A. Single 0.5-m interferometer

We will initially operate a single IFO with $L \approx 0.5$ m to rapidly test the design and to commission and characterize a high-power interferometer with the series of readout cavities and the SNSPD. This interferometer does not have the sensitivity to detect a quantum gravity signal due to its short arm length but will allow us to tackle the experimental challenges detailed in Appendix A. The goal of this experimental stage is to characterize the classical noise floor and achieve low photon count rates with an SNSPD in the absence of a detectable quantum gravity signal.

### B. Single 5-m interferometer

Having achieved sufficiently low noise levels in the 0.5-m interferometer phase, we will extend the arms of the interferometer, which increases the magnitude of the expected signal from geontropic fluctuations to a detectable level. This configuration will theoretically allow the GQuEST experiment to provide a significant detection of a quantum gravity signal using a single interferometer. The main observable of interest of the experiment is the average filtered photon flux out of the interferometer; the estimate thereof will be subject to a variance given by Poisson statistics, and this estimate, therefore, improves with increased measurement time [see Eq. (12)]. The identification of a signal in the average detected photon flux would first require the accurate subtraction of the noise level as characterized using the single 0.5-m interferometer (Sec. V A).

In case a significant excess photon flux is measured, follow-up investigations will be performed to determine if this signal is consistent with the expected signal from geontropic fluctuations. Specifically, the spectral shape of the signal can be measured by varying the filter offset frequency $\epsilon_r$. In addition, the dependence of the amplitude of the signal on the arm length of the IFO can be verified by changing the arm length. In the future, we could also vary the interarm angle $\Theta$ to verify the dependence of the signal on this parameter.

In case no significant excess photon flux is detected, i.e., if the average photon flux is consistent with the known noise, a constraint can be placed on the magnitude of the signal parametrized by $\alpha$.

### C. Twin interferometers

Once the experimental challenges have been addressed using a single IFO, a second identical IFO will be operated alongside the first. The detection of a geontropic signal through coherent signal detection of two interferometers does not require accurate characterization of the uncorrelated noise (see Sec. IV H). This is advantageous, as the uncertainty with which this noise level can be measured may be limited in practice, for example, due to irreducible uncertainty in the calibration or nonstationarity of the noise.

Under the assumption that the geontropic signal is largely correlated between collocated IFOs and the noise is uncorrelated, the measurement of a statistically significant nonzero correlated photon flux (which requires time averaging to reduce the measurement uncertainty) implies the presence of geontropic fluctuations, assuming there is no other coherent signal. Observation of such a correlated signal provides much stronger evidence than the observation of a signal in a single IFO, as the former is less likely to be spurious. Follow-up investigations of the same kind as for the single-interferometer stage can be performed to confirm the properties of the signal and rule out the possibility that the correlated photon flux is due to correlated noise (see Appendix A 12).

## VI. FUTURE DESIGN UPGRADES

### A. Signal recycling and filter improvements

Through the accelerated accrual of detection statistics that photon counting provides, and given the practical advantage of using coherent signal detection with twin IFOs, the fiducial design as specified above is sufficient for the goal of detecting geontropic fluctuations with $\alpha = 0.1$ at $3\sigma$ in a few months. However, with this design we are performing a narrow-band search for a wideband signal, thereby wasting over $(\Delta f - \Delta \epsilon)/\Delta f \approx 99.9\%$ of the signal power by rejecting it via the output filter cavities [compare Eq. (11) to Eq. (13)].

A potential upgrade to the design that would increase the signal power incident on the readout cavities is the implementation of signal recycling [47,48]. Signal recycling modifies the optical gain of the interferometer in a frequency-dependent way, i.e., $G \rightarrow G'(f)$. Signal recycling increases the optical gain (by a factor given by the finesse $\mathcal{F}_{SRC}$ of the signal-recycling cavity) within the bandwidth of that cavity and reduces the optical gain outside the bandwidth. As the signal power at the photodetector in the GQuEST design is limited by the narrow readout cavity bandwidth $\Delta \epsilon < \Delta f$, the signal-recycling bandwidth can encompass the whole readout frequency band and, thus, boost the signal sideband power by a factor of $\min\{\mathcal{F}_{SRC}, c/(2L\Delta\epsilon)\}$ [49]. The use of power recycling (as in the fiducial design) in addition to signal recycling is known as dual recycling, which has been demonstrated in other experiments [50,51], but presents operational challenges that we choose to avoid for the current GQuEST design.

Another possible design upgrade is the addition of more optical readout filter cavities in parallel to measure at more sideband frequencies simultaneously [34]. This allows for faster characterization of the signal and classical noise by resolving different parts of their spectra simultaneously; this is, therefore, an alternative to signal recycling. Emerging technology in quantum memories and optical signal processing devices could provide more efficient, simpler, and multiplexed narrow-linewidth optical filters to fully extract the broadband signal.

### B. Quantum state preparation

The quantum noise encountered in laser interferometers is due to the quantum state of the electromagnetic field entering the signal port of the interferometer [26]. The sensitivity of IFOs can be improved by quantum state preparation, by engineering and injecting optimized quantum states into the IFO instead of the default vacuum state.

An IFO that uses homodyne readout can be made more sensitive by preparing squeezed coherent states and injecting them into the IFO [26,52]; crucially, the reduction of quantum noise in this scheme is limited by optical losses [53,54]. In the realistic case that the reduction of quantum noise is limited by optical losses incurred along the signal path $\Lambda_{sig}$ (i.e., the optical path between the point of injection of the squeezed state and the detection of signals), the squeezed state injection reduces the quantum shot-noise level as $\bar{S}_L^q \rightarrow \Lambda_{sig}\bar{S}_L^q$. Thus, this improves the SNR$^2$ [see Eq. (7)] by a factor of $\Lambda_{sig}^{-2}$ and decreases the measurement time needed to detect a given signal by the same factor [55]. The observed squeezing level in current gravitational-wave IFOs (which use homodyne readout and injection of squeezed states) is at most 6 dB [56,57], limited primarily but not entirely by optical losses $\Lambda_{sig} = \mathcal{O}(20\%)$. This demonstrated squeezing level would enable an IFO using homodyne readout to detect geontropic fluctuations in a total measurement time that is a factor of $\mathcal{O}(20)$ shorter than the time it takes without squeezing. Future IFOs plan to use homodyne readout with squeezing levels of as much as 10 dB; this would decrease the required measurement time by a factor of 100, but implementing this level of squeezing requires extremely low optical losses.

As established in Sec. III, photon-counting readout fundamentally outperforms homodyne readout, yielding a reduction of measurement times by a factor of 70 for a realistic design and by a factor of $\mathcal{O}(10^7)$, in principle, limited by the anticipated classical noise floor [cf. Eq. (7) vs Eq. (12), and Eq. (19)]. However, this comparison does not account for quantum state preparation. We note that photon-counting readout still outperforms homodyne readout using state-of-the-art levels of squeezing. Moreover, photon-counting readout can also be improved through quantum state preparation, as shown in Ref. [30]; we review these findings in Appendix B.

In the case of photon-counting readout, the injection of squeezed states is not advantageous. Squeezed states have a Poisson distribution of even-number photon occupation and decohere irreversibly into thermal states from any amount of optical loss; squeezed states, therefore, inevitably produce additional background noise photons. However, injection of prepared Fock states can increase the SNR$^2$ by a factor of $(e\Lambda_{sig})^{-1}$ (where $e$ is Euler's number) [30]. To achieve this improvement, Fock states with an occupation of $N_{prep}$ photons need to be prepared in a temporal mode that is efficiently passed through the filter cavities; i.e., the states must be prepared to have a bandwidth of $\Delta\epsilon$ or smaller. In addition, the photon detector must be able to resolve the number of incident photons and detect states with an occupation of $N_{\text{obs}} \geq N_{\text{prep}} + 1$ photons with a low background rate. For comparison, in the absence of quantum state preparation (as in the current fiducial design), the default vacuum states, for which $N_{\text{prep}} = 0$, are implicitly injected and the superconducting nanowire single-photon detectors detect any state for which $N_{\text{obs}} \geq 1$. We note that quantum state preparation of specific states that saturate an extended-channel quantum information bound can accelerate stochastic signal searches even more than when using Fock states; the optimal states improve the SNR [cf. Eq. (12)] by a factor of $\Lambda_{\text{sig}}^{-1}$ [30].

## VII. SUMMARY AND OUTLOOK

Geontropic space-time fluctuations would manifest in the output of an interferometer as a broadband signal at angular frequencies on the order of the light-crossing frequency $c/L$. Photon-counting readout of an interferometer allows bypassing the standard quantum limit of interferometry at these frequencies, enabling an accelerated search for signals from geontropic fluctuations. The GQuEST experiment will implement this readout design for the first time and is expected to be limited by classical thermal noise from the interferometer optics. The experiment is projected to reach the nominal predicted geontropic signal PSD peak of $\bar{S}_L^\phi = (3 \times 10^{-22} \text{ m}/\sqrt{\text{Hz}})^2$ within several hours of integrated measurement time.

The detection of quantum space-time fluctuations of this magnitude would constitute the first evidence of the quantum nature of gravity. Such a detection would demonstrate two facets of quantum gravity: first, how gravity is quantized with a minimum uncertainty set by the Planck scale [8,9] and, second, how such quantum fluctuations must accumulate across a light-crossing time of a causal diamond in a holographic theory of quantum gravity [6,14]. While a UV-complete theory is not yet available, the low-energy effective theory as given by the pixellon model can readily be tested. The predicted signal from this model, the magnitude of which is parametrized by $\alpha$, can be unequivocally detected or constrained. In the case of nondetection, the constraint set on $\alpha$ provides a concrete guide for theoretical efforts in quantum gravity. Depending on the stringency of the constraint and the theoretical predictions currently being prepared, the experimental data could rule out geontropic space-time fluctuations entirely.

The successful demonstration of photon-counting readout of a laser interferometer would pave the way for future interferometry experiments to pursue this technique as a means to significantly increase their sensitivity. In particular, photon-counting readout has the potential to greatly exceed the sensitivity gain offered by quantum squeezing in a homodyne readout scheme, the only other known method to go beyond quantum limits. Thus, photon counting has the potential to profoundly improve the sensitivity of all laser interferometers performing searches of signals with random components or stochastic signals defined by a spectral density. This includes those that aim to detect gravitational waves [28,58,59] and dark matter [60–63].

## ACKNOWLEDGMENTS

The authors thank H. Siegel and Y. Levin for constructive comments on a previous version of the manuscript, highlighting the additional contribution of substrate thermorefractive noise due to standing optical waves in the beam splitter, and correcting factors of 2 in our charge carrier noise power. S. M. V. thanks James W. Gardner for discussions on the theory of quantum measurement of stochastic signals. This article was prepared by the GQuEST Collaboration using the resources of the Fermi National Accelerator Laboratory (Fermilab), a U.S. Department of Energy, Office of Science, Office of High Energy Physics HEP User Facility. Fermilab is managed by Fermi Research Alliance, LLC (FRA), acting under Contract No. DE-AC02-07CH11359. The GQuEST project is funded in part by the Heising-Simons Foundation through Grant No. 2022-3341.

## APPENDIX A: EXPERIMENTAL CHALLENGES AND DETAILED NOISE BUDGET

In this appendix, we expand on the expected experimental challenges that must be overcome to achieve the nominal IFO design with the sensitivity presented above. These challenges are primarily related to maximizing the circulating power, increasing the number of signal photons, and minimizing nonsignal light incident on the output photodetectors. We explore the specific physical effects that degrade photon-counting interferometer performance. The prominence of these effects depends in part on the materials used in the interferometer. Table II contains fiducial parameters for the experimental design. Power spectral densities of the various noise contributions are plotted in Fig. 5.

### 1. Laser noise

While laser noise is often characterized by a linewidth, a complete description requires considering the laser noise PSDs of relative intensity noise and phase noise. Both laser noise spectra imply the presence of photons at frequencies offset from the carrier, potentially creating noise in the passband of the filter cavities. For the laser system used, amplified spontaneous emission causes both noises to have a broadband spectrally white contribution that must be suppressed or removed to prevent the signal from being obscured.

TABLE II. Additional parameters of the fiducial IFO design. Material parameters are evaluated at room temperature.

| Parameter | Symbol | Value |
|---|---|---|
| Laser angular wave number | $k$ | $4 \times 10^6$ m$^{-1}$ |
| Input laser white phase noise PSD | $\bar{S}_{\text{in}}^{\text{LP}}$ | $(10^{-7}\text{ rad}/\sqrt{\text{Hz}})^2$ |
| Nominal filter offset frequency/readout frequency | $\epsilon_{\text{r}}$ | 17.6 MHz |
| Minimum practical filter offset frequency | $\epsilon_{\text{r}}^{\min}$ | 8 MHz |
| Maximum practical filter offset frequency | $\epsilon_{\text{r}}^{\max}$ | 40 MHz |
| End mirror reflectivity | $R_{\text{EM}}$ | $\geq 0.9999$ |
| Power-recycling mirror transmissivity | $T_{\text{PR}}$ | 500 ppm |
| Total (round-trip) fractional PR cavity loss | $\Lambda_{\text{tot}}$ | $\mathcal{O}(100)$ ppm |
| End mirror $1/e^2$ ($2\sigma$) intensity beam radius | $w$ | 3 mm |
| End mirror diameter | $d$ | 25.4 mm |
| End mirror thickness | $h$ | 2 mm |
| End mirror substrate material | c-Si | 294K crystalline Si |
| Beam splitter $1/e^2$ ($2\sigma$) intensity beam radius | $w$ | 3 mm |
| Beam splitter diameter | $d_{\text{BS}}$ | 38.1 mm |
| Beam splitter thickness | $h$ | 2 mm |
| Beam splitter substrate material | c-Si | 294K crystalline Si |
| c-Si density | $\rho_{\text{s}}$ | 2329 kg m$^{-3}$ |
| c-Si Young's modulus | $E_{\text{s}}$ | 156 GPa |
| c-Si Poisson ratio | $v_{\text{s}}$ | 0.265 |
| c-Si body wave quality factor (at $\epsilon_{\text{r}}$) | $Q_{\text{s}}$ | $\mathcal{O}(10^6)$ |
| c-Si thermal conductivity | $\kappa_{\text{s}}$ | 380 W m$^{-1}$ K$^{-1}$ |
| c-Si specific heat | $C_{\text{s}}$ | 710 J kg$^{-1}$ K$^{-1}$ |
| c-Si thermorefractive coefficient $\partial n/\partial T$ at $\lambda$ | $\beta_{\text{s}}$ | $2 \times 10^{-4}$ K$^{-1}$ |
| c-Si coefficient of thermal expansion | $\alpha_{\text{s}}$ | $2.5 \times 10^{-6}$ K$^{-1}$ |
| c-Si Index of refraction at $\lambda$ | $n$ | 3.48 |
| c-Si diffusion constant | D | $3.76 \times 10^{-3}$ m |
| c-Si Debye length | $\lambda_{\text{D}}$ | $4.33 \times 10^{-7}$ m |
| c-Si mean carrier density | $N_0$ | $< 10^{18}$ m$^{-3}$ |
| c-Si optical absorption coefficient | $\alpha_{\text{e}}$ | $1.2 \times 10^{-26}$ m$^{-3}$ |
| c-Si fractional power absorption at $\lambda$ | $\Lambda_{\text{Si}}$ | $2 \times 10^{-4}$ m$^{-1}$ |
| Fractional BS coating power absorption (assumed) | $\Lambda_{\text{c}}$ | 3 ppm |
| Fractional BS substrate power absorption | $\Lambda_{\text{s}}$ | 0.4 ppm |
| Coating material | ⋯ | Ta$_2$O$_5$ − SiO$_2$ |
| Coating thickness | $h_{\text{c}}$ | $\mathcal{O}(10)$ μm |
| Ta$_2$O$_5$ Young's modulus | $E_{\text{Ta}}$ | 120 GPa |
| SiO$_2$ Young's modulus | $E_{\text{SiO}_2}$ | 70 GPa |
| Ta$_2$O$_5$ Poisson ratio | $\nu_{\text{Ta}}$ | 0.29 |
| SiO$_2$ Poisson ratio | $\nu_{\text{SiO}_2}$ | 0.19 |
| Coating body wave quality factor (at $\epsilon_{\text{r}}$) (derived) | $Q_{\text{c}}$ | 1400 |
| Coating thermal conductivity (average) | $\kappa_{\text{c}}$ | 2.6 W m$^{-1}$ K$^{-1}$ |
| Coating density (average) | $\rho_{\text{c}}$ | 5200 kg m$^{-3}$ |
| Coating specific heat (average) | $C_{\text{c}}$ | 360 J kg$^{-1}$ K$^{-1}$ |
| Coating effective coefficient of thermal expansion | $\bar{\alpha}_{\text{c}}$ | $6 \times 10^{-6}$ K$^{-1}$ |
| Coating effective thermorefractive coefficient | $\bar{\beta}_{\text{c}}$ | $8 \times 10^{-6}$ K$^{-1}$ |
| c-Si effective coefficient of thermal expansion | $\bar{\alpha}_{\text{s}}$ | $6.4 \times 10^{-6}$ K$^{-1}$ |
| Coating stress | $\sigma_{\text{c}}$ | 0.5 GPa |
| Fused silica thermal conductivity | $\kappa_{\text{FS}}$ | 1.38 W m$^{-1}$ K$^{-1}$ |
| Fused silica thermorefractive coefficient at $\lambda$ | $\beta_{\text{FS}}$ | $8.5 \times 10^{-6}$ K$^{-1}$ |
| Fused silica fractional power absorption at $\lambda$ | $\Lambda_{\text{FS}}$ | $10^{-4}$ m$^{-1}$ |

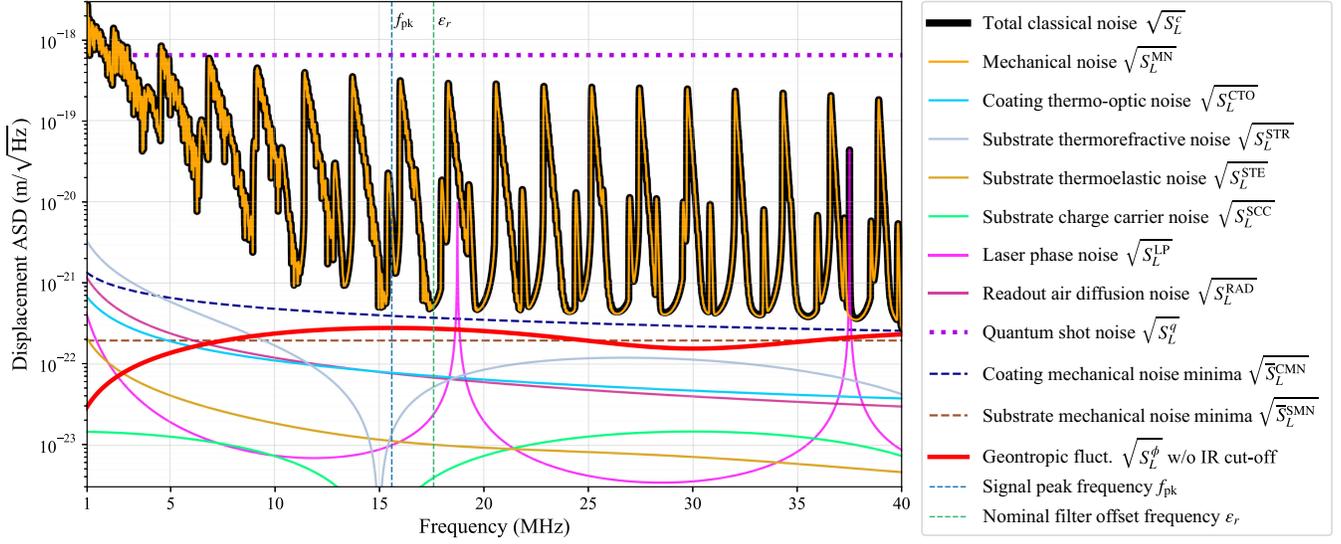

FIG. 5. The displacement amplitude spectral density of various noises as estimated for the fiducial design (see Table II) are plotted together with the expected signal from geontropic quantum space-time fluctuations $\sqrt{S_L^\phi}$ (red line). The total classical noise $\sqrt{S_L^c}$, which limits the sensitivity of GQuEST, and its contributions are shown (orange, blue, gray, gold, lime green, pink, and violet lines). The standard quantum limit from shot noise $\sqrt{S_L^q}$, which GQuEST is not subject to thanks to its photon-counting readout, is also shown (purple dotted line). The contributions of the coating and the substrate to the total mechanical noise $S_L^{\rm MN}$ at local minima are plotted as the curves $\sqrt{\overline{S_L^{\rm CMN}}}$ (dashed blue line) and $\sqrt{\overline{S_L^{\rm SMN}}}$ (dashed brown line), respectively (see Appendix A 4). The dark blue vertical dashed line marks the peak signal frequency $f_{\rm pk}$. The green vertical dashed line marks the fiducial cavity offset frequency $\epsilon_r$; note that the filter bandwidth $\Delta\epsilon = 25$ kHz is narrower than the thickness of that line in this plot.

This input laser white phase noise PSD is $S_{\rm in}^{\rm LP} = (\mathcal{O}(10^{-7})\ {\rm rad}/\sqrt{\rm Hz})^2$ (Table II). This amounts to a photon flux spectral density of

$$\bar{S}_{\dot{N}}^{\rm LP} \approx 4 \times 10^3 \left(\frac{S_{\rm in}^{\rm LP}}{\left(10^{-7}\ {\rm rad}/\sqrt{\rm Hz}\right)^2}\right)\left(\frac{P_\nu}{100\ {\rm mW}}\right)\left(\frac{0.8\ {\rm eV}}{h\nu}\right) \quad (A1)$$

at all optical frequencies, where $P_\nu$ is the power of the light at the laser frequency $\nu$. Given an output power of $P_\nu = P_{\rm out} = \mathcal{O}(100)$ mW and an output filter cavity bandwidth $\Delta\epsilon \approx 25$ kHz, this noise amounts to a photon flux at the photodetector of

$$\dot{N}_{\rm pass}^{\rm LP} = \int_{-\infty}^{\infty} S_{\dot{N}}^{\rm LP}(\epsilon) F(\epsilon - \epsilon_r) d\epsilon$$

$$\approx \Delta\epsilon \bar{S}_{\dot{N}}^{\rm LP} \approx 1 \times 10^8\ {\rm Hz}\left(\frac{P_{\rm out}}{100\ {\rm mW}}\right). \quad (A2)$$

This calculation assumes that the two interferometer arm lengths are exactly equal (i.e., no Schnupp asymmetry, $\Delta L_{\rm Schnupp} = L_1 - L_2 = 0$). A nonzero $\Delta L_{\rm Schnupp}$ can couple laser noise into the output port even on a perfect dark fringe; this effect is subdominant to the coupling through the contrast defect as given above [Eq. (A2)] as long as $\Delta L_{\rm Schnupp} < c/(2\epsilon_r)\sqrt{P_{\rm out}/P_{\rm BS}} \approx 3$ cm [64].

To combat this potentially large flux of amplified spontaneous emission photons [Eq. (A2)], the seed laser and amplifier are followed by a single passive input filter cavity. This cavity has a round-trip length of 4.5 m and mirrors with a transmissivity of 900 ppm, providing $\mathcal{O}(10^6)$ suppression of the phase noise power at frequencies a factor of $10^3$ above the cavity pole of $\mathcal{O}(10)$ kHz. The power recycling of the IFO has a similar cavity pole, and together the two cavities thus suppress the phase noise by up to a factor of $10^{12}$ in power, therefore reducing the laser phase noise to a level an order of magnitude below the signal photon flux. The laser phase noise is not suppressed at multiples of the free spectral range (FSR) of these cavities, which produces peaks in the noise spectrum as shown in Fig. 5.

### 2. Optical power recycling

We plan to have $\mathcal{O}(10)$ kW of circulating optical power in the IFO to maximize the signal photon flux [and, hence, the SNR; see Eq. (19)]. Maximizing the circulating power requires the minimization of optical losses inside the interferometer. Absorption or scattering inside the IFO, transmission of light through the end mirrors, and light leaving the IFO through the output port all limit the buildup of circulating power. Specifically, the circulating optical power in a power-recycling (PR) optical cavity as in Fig. 2 is [20]

$$P_{\rm BS} = G_{\rm PR} P_{\rm in}, \qquad G_{\rm PR} \equiv \frac{T_{\rm PR}}{\left[1 - \sqrt{R_{\rm PR} R_{\rm EM}(1 - \Lambda_{\rm tot})}\right]^2}, \quad (A3)$$

where $G_{\rm PR}$ is the gain of power-recycling cavity and the other variables and their fiducial values are defined in Table II. These parameters allow the desired circulating optical power $P_{\rm BS} = 10$ kW to be achieved with losses as high as $\Lambda_{\rm tot} \lesssim 800$ ppm, where the losses are the limiting factor for the power buildup if $\Lambda_{\rm tot} > T_{\rm PR}$. However, the experimental design aims to limit the total losses to $\mathcal{O}(100)$ ppm to provide a margin for obtaining the desired circulating power.

We note that for the traveling-wave power-recycling cavity configuration discussed in Appendix A 5 (which greatly reduces substrate thermorefractive noise) additional loss from the power-recycling cavity is expected. This additional loss occurs due to optical power being transferred into the reverse-propagating optical field, which has the same resonance frequency as the main forward-propagating field in a traveling-wave configuration. If a fraction $T_{\rm rev}^{\rm sc}$ of the forward-circulating power $P_{\rm BS}$ is scattered into the reverse-circulating field, this light is resonantly enhanced in the power-recycling cavity and partially leaks out the PRM. The amount of power transferred into the reverse-propagating field and, thus, lost from the forward-propagating field is

$$P_{\rm rev}^{\rm out} = T_{\rm rev}^{\rm sc} G_{\rm PR} P_{\rm BS} = T_{\rm rev}^{\rm sc} G_{\rm PR}^2 P_{\rm in}. \quad (A4)$$

Consequently, a fraction $P_{\rm rev}/P_{\rm in} = \Lambda_{\rm rev}$ of the input power is lost from the cavity every round trip; this loss contributes to the total power loss $\Lambda_{\rm tot}$.

To quantify this contribution $\Lambda_{\rm rev}$, we proceed to estimate $T_{\rm rev}^{\rm sc}$. For this, we assume the fraction of light that is scattered from an optic is $\Lambda_{\rm scatter} = \mathcal{O}(10)$ ppm. We further assume all this light is scattered into a cone with a half-angle equal to the angle of incidence (determined by the required separation of the incident and reflected beam) $\theta_i \approx 5 \times 10^{-3}$ rad, which likely is an overestimate of the amount of light scattered at these angles; the solid angle into which light is scattered is then $\Phi_{\rm cone} \lesssim 3 \times 10^{-4}$ sr. The resonant reverse-propagating optical mode with radius $w$ has a beam divergence angle $\theta_{\rm div} = 2/(kw)$ and subtends a solid angle $\Phi_w = \pi\theta_{\rm div}^2/2 = \mathcal{O}(10^{-8})$ sr from a scattering optic. Thus, we estimate the amount of light scattered into the reverse-propagating mode per round trip is $T_{\rm rev}^{\rm sc} \lesssim \Lambda_{\rm scatter} \Phi_w/\Phi_{\rm cone} < \mathcal{O}(10^{-9})$, which gives $\Lambda_{\rm rev} \lesssim \mathcal{O}(10^{-3}) = \mathcal{O}(T_{\rm PR})$. Thus, the scatter into reverse-propagating light could limit the power buildup [Eq. (A3)] in a traveling-wave power-recycling configuration; we leave a more precise analysis for future work.

### 3. Control of mirror positions

Another important part of the IFO design is the implementation of feedback control to maintain the IFO at the operating point. The position of the mirrors directly influences the amount of power circulating inside the interferometer; the length of the power-recycling cavity must be controlled to be resonant with the laser light to maintain the high-power circulating field. Deviations of the positions of mirrors cause power to be lost from the cavity. Moreover, differential fluctuations of the positions of the arm end mirrors produce noise light at the output.

Environmental noise couples to and perturbs the positions of the optics. Feedback control of the differential position of the end mirrors are implemented to counteract these perturbations. Homodyne readout can be used to measure this degree of freedom, and a homodyne readout channel is implemented using the residual fringe light power at the output $P_{\rm out}$ that is reflected from the first of the filter cavities (see Fig. 2, bottom left). This readout can alternatively be implemented with the balanced-homodyne scheme [65] to minimize the fringe light required to detect the length perturbation. Deviations of the mirror positions thus inferred are fed back to the mirror positions in a control loop, where the mirrors are actuated using piezo-electric transducers.

Given a target optical loss of $\mathcal{O}(10)$ ppm from motion, we estimate [using Eq. (8)] that the maximum allowable rms differential arm length deviation is $\Delta \langle \delta L_{12} \rangle = \mathcal{O}(1)$ nm. The end mirror actuator design is intended to achieve feedback control with a bandwidth of $O(700)$ Hz, similar to the performance of the Holometer [19].

### 4. Mechanical thermal noise in the optics

Thermal excitation of the optical components is expected to be a significant source of noise for the GQuEST experiment. The dominant noise source is elastic mechanical vibrations of the disk-shaped mirror substrates and of the optical coatings, known as solid normal modes (SNMs) as shown in Fig. 5. In general, thermal dissipation produces fluctuations in the optics that affect the optical path length of light interacting with the optic. This section treats thermal mechanical fluctuations (i.e., vibrations). This noise source has in previous literature been referred to as "Brownian thermal noise" [43,66,67]; we avoid the term "Brownian" since the underlying physical process does not involve mass diffusion. The following section (Appendix A 5) considers inhomogeneous dissipation, which produces temperature fluctuations inside the optic that results in noise on the incident light.

We consider two methods of modeling noise from homogeneous mechanical thermal excitation of the optics.

Both methods invoke the fluctuation-dissipation theorem (FDT) of Callen and Welton [68] to find the mechanical noise fluctuations conjugate to thermal dissipation. The first is the "direct" method as posited by Levin, which derives the noise by considering the power dissipated by an oscillatory force on the optic. The second method derives the noise by decomposing the excitation of the optic into normal modes and then considering these modes to be thermally populated according to the equipartition theorem.

The direct method proposed by Levin [69] is conventionally used in the interferometry community to model the PSD of mechanical thermal noise in the optics. In this method, one starts by considering an oscillatory force $F = F_{\text{pk}} \cos(\Omega t)$ applied to the optic surface, conceptually arising from the radiation pressure of an incident light beam. Here, $\Omega$ is the angular frequency at which the noise is to be computed (i.e., $\Omega = 2\pi f$ or $\Omega = 2\pi\epsilon$ in the sideband picture). The force produces elastic deformations of the material, which modulates the optical path length of the incident beam. This deformation stores elastic energy $U_{\max} \propto F_{\text{pk}}$ in the optic, which then partially dissipates when the material relaxes. The dissipated power is $W_{\text{diss}} = \Omega U_{\max}/Q$, where $Q(\Omega)$ is the quality factor of the mechanical system, which quantifies the internal damping of the material. The FDT implies that the thermal noise in a certain physical degree of freedom is determined by the dissipation occurring in response to a generalized force acting on that degree of freedom. Formally, the dissipation is quantified by the resistive part of the frequency response, i.e., the real part of the admittance $Y(\Omega)$ of the system.

Levin's direct method, thus, relates the force and the dissipated energy to compute the PSD of the fluctuating deformations of the optic produced by thermal energy in the material [69]:

$$S_L^{\text{FDT}}(\Omega) = \frac{4k_\text{B}T}{\Omega} \frac{U_{\max}}{QF_{\text{pk}}^2} = \frac{4k_\text{B}T}{\Omega^2} |\text{Re}[Y(\Omega)]|, \quad \text{(A5)}$$

where the parameters are defined above or in Table II. The mechanical frequency response of the optic, when driven by radiation pressure along a Gaussian beam profile, thus provides the needed information to compute the mechanical thermal noises. The conventional analytical application of Levin's method treats the optic as an infinite half-space of material, in which case there are no SNM resonances and the only physical scale is given by the beam width. This approach is, therefore, not applicable at frequencies comparable to SNM resonances, i.e., for the GQuEST measurement band. However, we use Levin's method to calculate coating contributions to the mechanical noise, as explained in Appendix A 4 b.

The optic's mechanical thermal noise is modeled by decomposing the mechanical excitation into normal modes of the optic, following Gillespie and Raab [70]. The noise spectrum is then given by a thermal population of these modes according to the equipartition theorem. The expected mechanical noise from SNMs, which we call the mechanical noise $S_L^{\text{MN}}(\Omega)$, as shown in Fig. 5 is modeled using this method, as we explicate below in Appendix A 4 a. Importantly, we find that the noise at frequencies between SNM resonances, i.e., the total noise $S_L^c(\epsilon_r)$, is mostly due to the effect of the optical coatings, and this contribution is quantified in Appendix A 4 b.

### a. Elastic solid normal modes in the optics

SNMs are the resonances of elastic body waves in the optical substrate materials. To analytically model the noise from SNMs in the optics, we start by considering the power spectral density of the displacement noise imparted on the light by a single mechanical resonance peak. We consider a complete set of orthogonal normal mechanical modes (SNMs) identified by their eigenfrequency $\{\omega_\kappa\}$, where $\kappa$ runs over all modes. These SNMs are obtained by solving the elastic wave equation. The noise PSD from a single SNM is [70]

$$S_L^{\text{MN}_\kappa}(\Omega) = C_\kappa \frac{4k_\text{B}T}{m\Omega(\omega_\kappa)^2} \left( \frac{Q_\kappa}{1 + Q_\kappa^2((\Omega/\omega_\kappa)^2 - 1)^2} \right), \quad \text{(A6)}$$

where $C_\kappa$ is a dimensionless parameter that describes the coupling of a mode to the incident light beam. It is related to the analogous mass-scale parameter $\alpha_n$ defined in Ref. [70] as $C_\kappa = 1/\alpha_n$, but we use a more convenient convention for summations. The mechanical quality factor $Q_\kappa$ of the mode $\kappa$ corresponds to the dissipative loss of the mode and can be decomposed as

$$\frac{1}{Q_\kappa} = \sum_i \frac{U_{\kappa,i}}{U_\kappa^{\max}} \varphi_{\kappa,i}, \quad \text{(A7)}$$

where the sum runs over the different parts or layers of the optic, i.e., the coating layers and the substrate; $\varphi_{\kappa,i}$ and $U_{\kappa,i}$ are the effective loss angle and elastic energy of the $i$th part for the mode $\kappa$, respectively, and $U_\kappa^{\max}$ is the total elastic energy in the mode $\kappa$. The other parameters in the equation are defined in Table II. We also refer to the contributions of the optical coating and the substrate to the quality factors as $Q_c$ and $Q_s$, which represent parts of the sum in Eq. (A7), i.e., the sum over either just the coating or the substrate, respectively. Note that evaluating Eq. (A7) requires evaluation of the fractional energies stored in the substrate $U_{\kappa,s}$ and the coatings $U_{\kappa,c}$, as well as knowledge of the effective loss angles in each part ($\varphi_{\kappa,s}, \varphi_{\kappa,c}$). While $U_{\kappa,s}$, $\varphi_{\kappa,s}$, and $\varphi_{\kappa,c}$ can be measured or obtained from solid mechanics theory, the evaluation of $U_{\kappa,c}$ requires a different treatment, as we explain in Appendix A 4 b.

The total displacement noise due to SNMs in the optic is

$$S_L^{MN}(\Omega) = \sum_\kappa S_L^{MN_\kappa}(\Omega). \quad (A8)$$

To identify all the modes and evaluate this sum, we use a Helmholtz decomposition of the elastic wave equations. This separates the modes into longitudinal (pressure) $P$-wave and transverse (shear) $S$-wave terms, which each have a different stiffness given by the $P$-wave modulus $M_s$ and $S$-wave modulus $G_s$, respectively. Note that the $P$-wave modulus $M_s = E_s(1 - v_s)/(1 - v_s - 2v_s^2)$ is the stiffness of purely longitudinal (axial) deformation, which is a different quantity than both the Young's modulus $E_s$ and the bulk modulus. The shear modulus is $G_s = E_s/(2(1 + v_s))$. The different elastic moduli of the two wave types lead to different wave propagation speeds, so these waves resonate at frequencies with different spacings such that the resonances do not overlap at high frequencies. This also means the $S$- and $P$-wave resonances do not strongly couple at the boundaries, which allows the $P$-wave and $S$-wave modes to be treated independently, simplifying the mode decomposition. The decomposition of the substrate's normal modes has been performed analytically for a square-shaped mirror, and the quadrature sum of noise from both end mirrors and the beam splitter yields the curve $S_L^{MN}$ as plotted in Fig. 5. We note that high-$Q$ modes expressed by Eq. (A6) may not be resolved given the finite frequency resolution of Fig. 5, but we integrate the average spectral density over each frequency bin so the plot accurately indicates rms noise density.

The analysis above predicts that the mechanical noise spectrum will be dominated by regularly spaced peaks corresponding to the mechanical resonances (SNMs) with a wave vector that is mostly parallel to the incident beam. These resonance peaks occur at frequencies $\Omega = n\omega_\kappa = n\pi v_s/h, n \in \mathbb{Z}$, where $v_s$ is the phonon propagation speed and $h$ is the longitudinal dimension of the optic (parallel to the beam, i.e., the thickness). $v_s = \sqrt{M_s/\rho_s}$ for $P$ waves and $v_s = \sqrt{G_s/\rho_s}$ for $S$ waves. The lowest-order SNMs have frequencies that are well below the measurement band of the GQuEST experiment. The density of SNM modes in frequency space is expected to grow quadratically with frequency. Therefore, as the modes are thermally populated according to the equipartition theorem, one might expect the total noise to grow at high frequencies. However, the noise is reduced because the coupling to the incident Gaussian beam (parametrized by $C_\kappa$) of higher-order modes, i.e., those where the vector that defines the direction of oscillation has a large component perpendicular to the beam axis, scales as

$$C_\kappa \propto e^{-w^2 k_\perp^2/4}, \quad k_\perp^2 \equiv k^2 - (\mathbf{k} \cdot \mathbf{n})^2, \quad (A9)$$

where $\mathbf{k}$ is the SNM wave vector, $\mathbf{n}$ is the unit vector parallel to the beam axis, and $\mathbf{k}_\perp$ is the transverse wave vector. The $S$ waves, which are by definition transverse (shear) waves, produce motion of the reflecting surface only when their wave vector has a component perpendicular to the beam axis. Therefore, the contributions to the noise of $S$-wave resonances are reduced [by a factor of $\mathcal{O}(10^2)$ in noise power] compared to the $P$-wave modes. Subsidiary higher-order SNM peaks near a primary SNM resonance add noise on top of the noise floor (see Fig. 5). This contribution can be mitigated by using a large beam radius $w$, which reduces the coupling of modes with wave vectors perpendicular to the beam.

The strategy for GQuEST is to tune the output filter cavities such that a signal can be measured at frequencies between the $P$-wave SNM noise peaks, at the noise floor of $\bar{S}_L^c \approx S_L^{MN}(\Omega)$.

To motivate our choice of optic substrate material and thickness, we consider the level of the resulting noise floor between the mode peaks at the measurement readout frequency $\epsilon_r$ due to the properties of the substrate, factoring out the effect of coatings. We, thus, decompose the total mechanical noise as

$$S_L^{MN}(\Omega) \geq \bar{S}_L^{SMN} + \bar{S}_L^{CMN}(\Omega), \quad (A10)$$

where the superscripts indicate the respective contributions to the noise floor from substrate mechanical noise (SMN) and coating mechanical noise (CMN); the latter is later derived in Eq. (A14). Eq. (A10) holds as an approximate equality at local minima of $S_L^{MN}(\Omega)$ at higher frequencies, where the coupling of high transverse-wave-number modes (i.e., modes for which $w^2 \mathbf{k}_\perp^2 \gg 1$) is greatly reduced [see Eq. (A9)]. We express the contribution to the noise floor from mechanical dissipation in the substrates (parametrized by $Q_s = 1/\varphi_s$) using the following approximate analytical expression:

$$\bar{S}_L^{SMN}(\Omega) \approx \frac{16 k_B T h}{\pi^3 v_s^2 \rho_s w^2 Q_s \Omega} = \frac{16 k_B T h \varphi_s}{\pi^3 M_s w^2 \Omega}. \quad (A11)$$

This equation defines a curve that intersects the local minima of the SNM noise when the effect of coatings is neglected. We leave the derivation of this expression for future work but provide it to indicate the scaling of the noise floor with design parameters. To increase the frequency separation between mechanical resonances and lower the noise floor between them, a thinner and stiffer optic with a higher quality factor is desirable. The mechanical quality factor of the silicon substrate is expected to be limited by Akhiezer damping and is given by $Q_s \cdot \Omega = 3 \times 10^{15}$ Hz [71] for $Q_s$ as given in Table II evaluated at $\Omega = 2\pi\epsilon_r$. This frequency dependence of the quality factor makes the PSD [Eq. (A11)] flat, as seen in

Fig. 5. The fiducial design uses crystalline silicon optics with a thickness of $h = 2$ mm.

In addition to the computation of the noise from SNMs via the decomposition of the optic's excitation into normal modes, i.e., the evaluation of Eq. (A8), the expected thermal noise spectrum from SNMs has also been modeled with Levin's direct method. This was done numerically using COMSOL Multiphysics®, a finite element modeling program that performs the necessary volume integrals over energy and dissipation. Both modeling methods agree on the level of the noise floor $S_L^{\mathrm{MN}}(\epsilon_\mathrm{r}) \approx (10^{-21}\ \mathrm{m}/\sqrt{\mathrm{Hz}})^2 \approx S_L^\mathrm{c}(\epsilon_\mathrm{r})$. Moreover, the analytical model agrees with data from the Fermilab Holometer [19]. The model captures the characteristic "sawtooth" shape in the measured noise spectrum that arises from the density of SNM states and the coupling to the Gaussian beam mode.

Mechanical thermal noise also arises from the mirrors that compose the power-recycling cavity. This noise does not typically impact homodyne readout, as the noise is common between the arms and, therefore, appears only in the amplitude quadrature of the optical field at the output, whereas the signal is in the phase quadrature. However, photon counting measures signal and noise in both optical quadratures and is, thus, in principle, sensitive to common-mode noise, if this noise is coupled to the output through asymmetries intrinsic to the interferometer. However, common-mode thermal noise, like input laser noise, is suppressed by the power-recycling cavity and attenuated by operating the interferometer close to a dark fringe, making it subdominant to thermal noise in the end mirrors and beam splitter.

Finally, SNMs of the optics also scatter light from the fundamental optical mode into higher-order modes. We anticipate that this scattering into higher-order modes will not be negligible and will change the frequency dependence of the noise spectral density from higher-order SNMs. We expect that the readout filter cavities will fully suppress this effect as they reject higher-order optical modes but leave analysis of this effect for future work.

### b. Mechanical thermal noise from optical coatings

Equation (A8) and the corresponding curve in Fig. 5 fully incorporate the thermal noise from the optics, including the contribution added by optical coatings through Eq. (A7) in the form of $Q_\mathrm{c} = 1/\varphi_\mathrm{c}$. We evaluate this contribution separately in this section, as it requires a different approach to the evaluation of the contribution of the substrate $Q_\mathrm{s}$. We find the loss in the coatings contributes significantly to the noise floor, as represented by Eq. (A10), and we refer to this contribution as the CMN.

The multilayer Bragg-reflector coatings exhibit both mechanical (homogeneous) dissipation and thermal diffusion (inhomogeneous dissipation). Both forms of dissipation contribute noise to the optical beam. The homogeneous term relating to mechanical noise is considered here, while the latter term from thermal diffusion is called thermo-optic noise and treated in Appendix A 5 b.

Beyond the mechanical vibrations of the reflecting (front) surface of the optic, additional mechanical effects influence the light. Mechanical excitations inside the coating entail changes of the size and index of refraction of the layers of the Bragg-reflector coating, which, in turn, gives rise to fluctuations of the reflection phase of the incident optical field [43]. We do not consider the effect of reflection phase changes here, as it is a subdominant $\mathcal{O}(30\%)$ contribution in GW detectors [43,67].

The mechanical noise of the coating can be treated in unison with that of the substrate, by applying solid mechanics theory for a composite solid [72]. However, solving the homogeneous elastic wave equations for the compound optic does not provide the correct value for $U_{\kappa,\mathrm{c}}$, as the free boundary conditions imply no energy is stored in the coatings when a normal mode is excited and no significant noise is contributed by the coatings, which is contrary to experimental observation and alternative theoretical treatments.

To estimate the coating mechanical noise, the elastic wave equation must be solved with a force at the boundary that corresponds to the incident beam, whereby we utilize Levin's direct method while treating an optic of finite size. The solution of the wave equation provides the admittance $Y_\kappa$ of the optic surface when acted on by the force at the boundary, which now contains the total $Q_\kappa$ according to Eq. (A5). The energy stored in the coatings $U_{\kappa,\mathrm{c}}$ can then be obtained by factorization of Eq. (A7), which finally allows the total mechanical noise to be computed in Eq. (A8).

This factorization uses measured values of the mechanical loss angles of the coatings $\varphi_{\kappa,\mathrm{c}}$. The mechanical loss angles of the high-reflection optical coatings used in aLIGO mirrors have been inferred from direct measurements of the CMN at frequencies from 30 Hz to 2 kHz in Ref. [73]. These coatings are made of alternating layers of $\mathrm{SiO_2}$ and either $\mathrm{Ta_2O_5}$ or $\mathrm{TiO_2\!:\!Ta_2O_5}$. For $\mathrm{Ta_2O_5}$, the measurements give

$$\varphi_\mathrm{Ta} = (5.3 \pm 0.1) \times 10^{-4} \left(\frac{f}{100\ \mathrm{Hz}}\right)^{0.06 \pm 0.02}. \quad (\mathrm{A}12)$$

To estimate $\bar{S}_L^{\mathrm{CMN}}$ in the GQuEST experiment, we assume a coating design similar to aLIGO (i.e., the same number and thickness of alternating layers of $\mathrm{SiO_2}$ and $\mathrm{Ta_2O_5}$) with $\varphi_\mathrm{Ta}$ as given by Eq. (A12) and $\varphi_{\mathrm{SiO_2}} = 5 \times 10^{-5}$. We then extrapolate from the measurements in Ref. [73] and assume these values are accurate at frequencies in the band approximately 1–40 MHz. Specifically, we use

$$\varphi_{\kappa,\mathrm{c}} = \sum_{i=1}^{N_\mathrm{coat}} \varphi_{\mathrm{c},i} \frac{2h_{\mathrm{c},i} M_{\mathrm{c},i}}{h_\mathrm{c} M_\mathrm{s}}, \quad \text{where } \varphi_{\mathrm{c},i} = \begin{cases} \varphi_\mathrm{Ta} & i \text{ even} \\ \varphi_{\mathrm{SiO_2}} & i \text{ odd,} \end{cases}$$

$$(\mathrm{A}13)$$

where $h_{c,i}$ are the thicknesses of each coating layer, $h_c = \sum h_{c,i}$ is the total coating thickness, $\varphi_{c,i}$ are the loss angles of each layer, and $M_{c,i}$ is the P-wave modulus of the $i$th coating layer. When considering S waves, i.e., for certain $\kappa$, $M_{s,c}$ should be replaced with $G_{s,c}$ in this equation. These moduli can be calculated from the coating's Young's modulus and Poisson ratio using the same formula as for the substrate's moduli.

By considering $\varphi_c \approx \varphi_{\kappa,c}$ and including only the dominant P-wave contributions in the computation of $\varphi_{\kappa,c}$, we now evaluate the contribution of the coatings to the SNM thermal noise floor at the measurement frequency. When including only contributions from the coating to the total loss angle, we find the following analytical expression for the solid normal mode thermal noise floor:

$$\bar{S}_L^{CMN}(\Omega) \approx \frac{16 k_B T h_c \varphi_c}{\pi^3 M_s w^2 \Omega}. \quad (A14)$$

The remaining parameters are defined in Table II. This equation is analogous to Eq. (A11), except it uses the thickness and loss of the coating rather than those of the substrate. Together, Eqs. (A14) and (A11) can be used to evaluate Eq. (A7), to give an effective description of the two dominant dissipation contributions as $1/Q_\kappa \approx \varphi_s + (h_c/h)\varphi_c$. Note, however, that the substrate modulus $M_s$ appears in the denominator of both Eqs. (A14) and (A11) and in Eq. (A13) due to its dominant contribution to all SNM mode energies. The form of Eq. (A14) is comparable to the form of Eq. (1) in Ref. [67], though it has different weighting factors of the elastic moduli.

In conclusion, we find, thus, that the effective contributions to the total quality factor of the coating $Q_c \approx 1400$, compared to that of the substrate $Q_s \approx 10^6$, while the coating is roughly 200 times thinner than the substrate. Despite this, the coating contributes 3–4 times as much noise as the substrate.

### 5. Thermorefractive and thermoelastic noise in the optics

Thermorefractive noise and thermoelastic noise, collectively called thermo-optic noise, are due to random fluctuations in temperature in the optics from inhomogeneous dissipation. These temperature fluctuations produce corresponding changes in the index of refraction and the size of optical substrates and coatings. The changes are proportional to the materials' thermorefractive coefficients and coefficients of thermal expansion, respectively, and this produces phase noise in the incident beam.

#### a. Thermo-optic noise in optical substrates

The main contribution of thermo-optic noise in the optical substrates to the total noise in the IFO output is thermorefractive fluctuations in the beam splitter. The power spectral density of substrate thermorefractive (STR) noise is given by the following equation [44,74]:

$$S_L^{STR}(\Omega) = \frac{4 k_B \kappa_s T^2 \beta_s^2 h}{\pi (C_s \rho_s w^2 \Omega)^2 \cos(\theta_2)} \frac{\eta + \eta^{-1}}{2\eta^2}. \quad (A15)$$

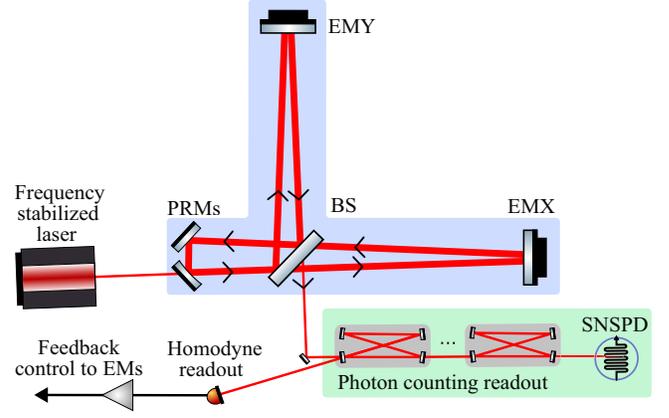

FIG. 6. Diagram of the design for a single GQuEST interferometer that includes a traveling-wave configuration for the power-recycling cavity. It is anticipated that this configuration is required to avoid having a standing optical wave in the beam splitter, which would produce significant additional thermorefractive noise [44] (see Appendix A 5).

$\theta_2$ is the angle of refraction in the beam splitter, and $\eta$ is defined in Ref. [44]; the parameter $\eta$ accounts for the elliptical cross section of the beam inside the beam splitter. In our estimate of the total STR, we do not include the contribution due to optical standing waves in the beam splitter, as described in Ref. [44]. This is because the final design of GQuEST will spatially separate the optical paths to and from the end mirrors such that no optical standing waves are formed and this noise contribution is avoided, as shown in Fig. 6. This standing wave noise term would increase the STR noise at our fiducial readout frequency by a factor of 36 in amplitude, greatly decreasing our sensitivity. For the fiducial design parameters of the GQuEST experiment, the thermorefractive noise amplitude spectral density (ASD) at the measurement frequency $\Omega/(2\pi) = \epsilon_r \approx 17.6$ MHz is, thus, $\sqrt{S_L^{STR}} = \mathcal{O}(10^{-22})$ m/$\sqrt{\text{Hz}}$. However, the measured noise at the interferometer output is modulated by the transfer function for phase modulations imparted at the beam splitter $H(\Omega) = \cos^2(\Omega L/c) \leq 1$. Thus, the total thermorefractive noise measured is $H(\Omega) S_L^{STR}(\Omega)$.

Thermoelastic noise arises from random fluctuations in temperature that result in the thermal expansion of all the optics (not just the beam splitter). The power spectral density of substrate thermoelastic (STE) noise in our measurement band is [75]

$$S_L^{\text{STE}}(\Omega) = \frac{8}{\sqrt{2\pi}} \frac{k_B \kappa_s T^2 \alpha_s^2 (1+v_s)^2}{C_s^2 \rho_s^2 w^3 \Omega^2}, \quad \text{(A16)}$$

where the variables are defined in Table II. For the reference design, the thermoelastic noise ASD at 17.6 MHz is $\sqrt{S_L^{\text{STE}}} = \mathcal{O}(10^{-23})$ m/$\sqrt{\text{Hz}}$. While the thermorefractive and thermoelastic noise in silicon is larger than in conventional fused silica optics, these noises are both well below the noise floor from solid normal modes. The thermoelastic noise from the beam splitter (but not that from the end mirrors) will be modulated by $H(\Omega)$.

### b. Thermo-optic noise in optical coatings

In this section, we consider the effects of thermoelastic and thermorefractive fluctuations in the optical coatings (CTE and CTR noise, respectively); we refer to their combined effect as coating thermo-optic (CTO) noise: $S_L^{\text{CTO}} = S_L^{\text{CTE}} + S_L^{\text{CTR}}$. The CTO noise $S_L^{\text{CTO}}(\Omega)$ is plotted in Fig. 5. We use the model in Ref. [76] as a starting point and follow their notation. However, their model describes noise at relatively low frequencies, as relevant in gravitational-wave detectors. Several key physical assumptions must, therefore, be reconsidered to model noise in the GQuEST measurement band. In particular, the coherence properties of this noise and its frequency dependence are different at higher frequencies, and this must be accounted for. Specifically, at low frequencies, the CTE and CTR contributions coherently cancel each other, while at high frequencies the noises are independent. In addition, at high frequencies, thermal fluctuations have coherence lengths that are shorter than the relevant physical coupling scale, which means the effect of the thermal fluctuations averages out to some degree, as explained below.

The PSDs of CTE and CTR noise are [76–78]

$$S_L^{\text{CTE}}(\Omega) = \frac{2\sqrt{2}k_B T^2 \Gamma_{\bar{\alpha}}(\Omega)}{\pi w^2 \sqrt{\kappa_s \rho_s C_s \Omega}} \left(\bar{\alpha}_c h_c - \bar{\alpha}_s h_c \frac{\rho_c C_c}{\rho_s C_s}\right)^2, \quad \text{(A17)}$$

$$S_L^{\text{CTR}}(\Omega) = \frac{2\sqrt{2}k_B T^2 \Gamma_{\bar{\beta}}(\Omega)}{\pi w^2 \sqrt{\kappa_s \rho_s C_s \Omega}} (\bar{\beta}_c \lambda)^2, \quad \text{(A18)}$$

where the overbars denote averaged effective coating material properties as defined in Ref. [76] and we introduce the dimensionless cutoff parameters $\Gamma_{\bar{\alpha}}(\Omega)$ and $\Gamma_{\bar{\beta}}(\Omega)$; the other variables are defined in Table II. These cutoff parameters parametrize the effect that thermal fluctuations with a scale smaller than the size of the part of the material that couples the fluctuations to the incident light are averaged out, and, thus, their effect diminishes. The scale of the thermal fluctuations is given by the thermal diffusion length $r_T \equiv \sqrt{\kappa_c/(\rho_c C_c \Omega)} \approx 0.1$ μm for

$\Omega = 2\pi \cdot 17.6$ MHz (see Ref. [76]). The scale of the thermoelastic coupling is the thickness of the coating $h_c$, as fluctuations throughout all layers contribute to the overall shift of the reflecting surface. The scale of the thermorefractive coupling is the depth that the optical field penetrates into the coating, $\bar{\lambda}$, given below. In the fiducial design $r_T < \bar{\lambda} \ll h_c$, which effects an additional roll-off rate of $1/\Omega$ and $1/\Omega^{3/2}$ in the CTE and CTR PSDs, respectively, in the measurement band and therefore greatly reduces these noises.

The asymptotic forms (i.e., at low and high frequencies) of the cutoff parameters for the CTE and CTR contributions, respectively, are

$$\Gamma_{\bar{\alpha}}(\Omega) = \frac{1}{\mathcal{O}(1 + R(1+R)h_c^2/r_T^2)}, \quad \text{(A19)}$$

$$\Gamma_{\bar{\beta}}(\Omega) = \frac{1}{\mathcal{O}\left(1 + \sqrt{2}R\bar{\lambda}^3/r_T^3\right)}, \quad \text{(A20)}$$

where $R = \sqrt{\kappa_c \rho_c C_c/(\kappa_s \rho_s C_s)}$, $\bar{\lambda}$ is defined below, and the other variables are defined above. The exact expression for $\Gamma_{\bar{\alpha}}$ is given in Ref. [76], where it is referred to as the thick coating correction, and is considered in more physical detail in Sec. IV.C.3 in Ref. [79]. This cutoff parameter causes coating thermo-optic noise to be dominated by the thermorefractive term.

The parameter $\Gamma_{\bar{\beta}}$ has not been considered in previous work; it gives the cutoff of the CTR and is a function of the depth that the optical field penetrates into the coating $\bar{\lambda}$. Specifically, we have that the intensity $I$ decays exponentially as a function of the distance traveled into the coating $z$ with the decay constant $\bar{\lambda}$:

$$I(z>0) \propto e^{-z/\bar{\lambda}}, \quad \bar{\lambda} = \frac{\lambda}{8\ln(n_H/n_L)}\left(\frac{1}{n_L} + \frac{1}{n_H}\right), \quad \text{(A21)}$$

where $n_H$ and $n_L$ are the indices of refraction of the high and low index of refraction coating layers, respectively. Previous work uses the approximation that the thermal diffusion length is much larger than the penetration depth of the beam into the coating, $r_T \gg \bar{\lambda}$ [cf. the delta function of Eq. (32) in Ref. [76] and Secs. V and VI in [80]], which is valid as those works model noise in the regime $r_T \gg \bar{\lambda}$. For the fiducial GQuEST design, $\bar{\lambda} \approx 0.62$ μm.

For the CTE and CTR noise PSDs plotted in Fig. 5, we use the exact form for $\Gamma_{\bar{\alpha}}$ from Ref. [76] and the approximate form for $\Gamma_{\bar{\beta}}$ given above, which we expect is accurate within a factor of 2.

The cutoff scales are important in modeling CTO noise, as they render what would otherwise be the dominant noise source at the measurement frequency subdominant. Future work will establish an exact form for $\Gamma_{\bar{\beta}}$ and should use a more precise description of the optical field in the coating, capturing the fact that the incident field is a standing wave that experiences discretized attenuation through successive layers. This description may add additional frequency scales corresponding to the thicknesses of individual layers ($\lambda/4$ or $\lambda/2$).

### 6. Charge carrier noise in the beam splitter

For semiconducting optics, another source of thermal noise inherent to the substrate needs to be considered. Thermal density fluctuations of the electrons in the conduction band of the material produce local fluctuations in the refractive index, which produces a noise with a PSD:

$$S_L^{\text{SCC}}(\Omega) = \frac{4Dk^2\alpha_e^2 N_0 h}{\pi w^2}\left[\frac{1}{\Omega^2 + (4Dk^2 + D/l_D^2)^2}\right]. \quad \text{(A22)}$$

Here, $\alpha_e = \partial n/\partial N$, where $N$ is the local number density of charge carriers (see Ref. [81], where $\alpha_e$ is called $\alpha$) and $N_0$ is the mean number density of charge carriers for the material, which is calculated in Ref. [82], where $N_0$ is called $n_0$. The other variables are defined in Table II.

For a silicon beam splitter at 294 K, as in the fiducial design, we expect the charge carrier noise to be $\sqrt{S_L^{\text{SCC}}} \leq \mathcal{O}(10^{-22})$ m/$\sqrt{\text{Hz}}$ based on the analysis in Ref. [81]. Similar to thermorefractive noise, this noise source is also modulated by the transfer function of phase modulations imparted at the beam splitter $H(\Omega)$. Practically, this calculation indicates that beam splitter substrates should have a high resistivity and correspondingly a low dopant concentration, in which case this noise contribution will be negligible.

### 7. Thermal lensing in the beam splitter

As the incident laser beam is partially absorbed by the coating and substrate of the beam splitter and because the beam has a nonconstant cross-sectional intensity profile, a thermal gradient is formed transverse to the optic axis. Because of the nonzero thermo-optic coefficient $\beta = \partial n/\partial T$, the transmitted light is lensed. This "thermal lens" effectively converts the incident fundamental-mode light into higher-order modes. As only the light transmitted through the beam splitter is lensed, the effect is differential and, thus, increases the contrast defect of the interferometer. Silicon has a higher thermal conductivity than fused silica, and, therefore, the thermal gradient and the resulting thermal lensing are reduced compared to high-power interferometers using a fused silica beam splitter.

The fraction of the contrast defect power output from the Michelson can be computed using the Laguerre-mode overlap integrals in Ref. [83] and accounting for the number of beam passes. This simplifies to the expression

$$\Lambda_{\text{defect}} = 0.07\eta\left(\frac{\beta_s}{\kappa_s \lambda}(\Lambda_c + \Lambda_s)P_{\text{BS}}\right)^2, \quad \text{(A23)}$$

where $\Lambda_{\text{defect}}$ is the fractional power loss into higher-order modes from the wavefront distortion. $\Lambda_s$, the power absorbed in the substrate, is equal to the thickness of the substrate times the absorption per unit length. The coefficient 0.07 is a geometrical factor that represents the sum of the squared inner products of all higher-order modes with the transverse profile of the thermal lens. The inner products are calculated using Eq. (7) in Ref. [83]. The other variables are defined in Table II.

The factor $\eta \approx 0.94\cos^2(2\phi_g) + 0.06$ incorporates the one-way Gouy phase advance $\phi_g$ of the beam going down an interferometer arm, assuming a beam waist either at the beam splitter or at the end mirror. We indicate the effect of $\eta$ to show that thermal lensing can be reduced in interferometers with a specific Gouy phase to cancel the contribution from the first-order Laguerre-Gauss modes to contrast defect light. GQuEST will use beams with large radii $w$ to suppress mechanical noises (see above), which entails that $\phi_g \ll 45°$ and $\eta \sim 1$.

The contrast defect power due to thermal lensing is then

$$P_{\text{ASdefect}} = \Lambda_{\text{defect}}P_{\text{BS}}. \quad \text{(A24)}$$

For GQuEST's design, $\Lambda_{\text{defect}} = 2$ ppm, and, therefore, we estimate that $P_{\text{ASdefect}} = 20$ mW. The effect of thermal lensing is proportional to the square of the substrate material property $\beta_M/\kappa_M$. For fused silica $(\beta_{\text{FS}}/\kappa_{\text{FS}})^2 = 3.6 \times 10^{-11}$ (m/W)$^2$, while for silicon $(\beta_s/\kappa_s)^2 = 1 \times 10^{-14}$ (m/W)$^2$; the choice of silicon, therefore, produces a reduction of thermal-lensing-induced contrast defect power by a factor of $\mathcal{O}(10^3)$.

For reference, using the same model to compute the expected thermal lensing for the case of the Fermilab Holometer [19], we find a value that agrees with the measured power lensed to within an order of magnitude. Therefore, the estimate for the thermal lensing in the GQuEST experiment is expected to be similarly accurate.

### 8. Optic curvature due to coating stress

High-reflectivity (HR) mirrors comprise dielectric (Bragg) reflection coatings applied to an optical substrate. This coating introduces elastic stress to the mirror surface, which changes the mirrors' radius of curvature. The effects of coating stress are expected to be greater for GQuEST than in other precision interferometers, as we use relatively thin optics to increase the frequency separation of mechanical resonances, and the coating stress, therefore, introduces more curvature than would occur for a thicker and, thus, stiffer mirror. The curvature induced in a flat mirror due to coating stress can be approximated in terms of its radius of curvature $r_{\text{curv}}$ and optical power $D = 2/r_{\text{curv}}$ using the Stoney equation [84], which gives

$$r_{\text{curv}} \approx \frac{1}{6}\left(\frac{E_s h^2}{\sigma_c h_c (1-v_s)}\right), \quad D \approx 12\left(\frac{\sigma_c h_c (1-v_s)}{E_s h^2}\right), \quad \text{(A25)}$$

where the variables are defined in Table II. Note that this approximation is valid for $(h_c/h) \ll 1$; for the GQuEST mirrors, $(h_c/h) = \mathcal{O}(10^{-2})$. Using the proposed experimental parameters, this gives an induced curvature of $r_{\text{curv}} = 7.6$ m or 0.26 diopters of spurious focusing power.

Differences in the induced curvature between the end mirrors would lead to a "mode mismatch" between the arms of the interferometer, which produces a contrast defect consisting of higher-order modes (HOMs). Specifically, a difference in the stress-induced curvature of the end mirrors along a direction $x$, $y$ orthogonal to the beam axis, i.e., $D_x = D_x^{\text{EMX}} - D_x^{\text{EMY}}$, scatters light from the fundamental mode into the (Hermite-Gauss) HG20 and HG02 modes with amplitude coefficients [85]

$$K_{20} \approx \frac{1}{\sqrt{2}}\left(\frac{kD_x w^2}{4}\right), \quad K_{02} \approx \frac{1}{\sqrt{2}}\left(\frac{kD_y w^2}{4}\right), \quad \text{(A26)}$$

where the parameters are in Table II. This scattering then gives rise to a contrast defect $\Lambda_{\text{CD}} = K_{02}^2 + K_{20}^2$. Thus, to achieve $\Lambda_{\text{CD}} < 10$ ppm, we require that the curvature mismatch between end mirrors $D_{\text{tot}} = \sqrt{D_x^2 + D_y^2} < 3 \times 10^{-4}$ diopters. This indicates that the coating-stress-induced curvature calculated above must somehow be compensated to satisfy this condition. We intend to partially compensate for the curvature induced by the HR coating by applying an AR coating with a custom thickness to the back of the optic, such that the stress induced by the AR coating cancels the curvature induced by the HR coating. However, this method requires *a priori* knowledge of the induced coating stress, and, as modeling of this coating stress will likely be accurate only to within approximately 1%–10%, the curvature can be compensated only to that fraction. The remaining residual differential curvature will be compensated using custom mirror mounts that actuate on the mirror such that its curvature can be corrected by an amount $\mathcal{O}(10)$ mD along two independent axes.

### 9. Optical cavities to form the narrow bandpass readout filter

The circulating power incident on the beam splitter of $P_{\text{BS}} = \mathcal{O}(10)$ kW corresponds to $\mathcal{O}(10^{23})$ photons/s. A contrast defect $\Lambda_{\text{CD}} = \mathcal{O}(10)$ ppm would amount to $P_{\text{out}} = \mathcal{O}(0.1 \text{ W} \approx 10^{18})$ photons/s at the interferometer output. Our goal is to suppress this light to achieve a photon flux smaller than that from the interferometer thermal noise [$\mathcal{O}(10^{-2})$ Hz] at the signal peak frequency. Therefore, in total, $\mathcal{O}(200)$ dB of power suppression of carrier laser light is needed. GQuEST's design includes four filtering cavities, each providing nearly 60 dB of suppression, to meet this goal.

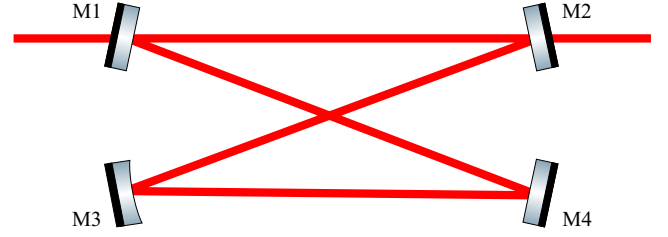

FIG. 7. The optical layout of the optical bow-tie filter readout cavities, comprised of four mirrors $M1 - M4$. Note that $M3$ is curved for cavity stability and design of the round-trip Gouy phase.

These filter cavities have a bow-tie geometry, with an optical path length of approximately 2.4 m, giving a FSR of roughly 125 MHz (see Fig. 7). The input and output couplers (mirrors) are given transmissivities of $T_I = T_O = 1000$ ppm, and, therefore, the finesse of each cavity is $\mathcal{F} \approx \pi/T_I = 3150$ and their bandwidth is $\Delta\epsilon_1 = 42$ kHz. Within the resonant bandwidth of each cavity, approximately 98% of the light is passed and 2% is lost, assuming 10 ppm of optical loss on each optic. Moreover, the filter cavity lengths will all be slightly different to give them slightly different FSRs, which prevents light at frequencies that are a multiple of the FSR from leaking through the filter cavities.

The filter function for a single output filter cavity is Lorentzian, and, therefore, the power attenuation of four cavities follows

$$F(\epsilon - \epsilon_r) = \left(\frac{\Delta\epsilon_1^2}{\Delta\epsilon_1^2 + 4(\epsilon - \epsilon_r)^2}\right)^4. \quad \text{(A27)}$$

Here, $\Delta\epsilon_1$ is the bandwidth of a single cavity. The integrated bandwidth $\Delta\epsilon = \int_{-\infty}^{-\infty} F(\epsilon)d\epsilon \approx \Delta\epsilon_1/2$. Based on this filter shape, $\epsilon_r$ cannot be set arbitrarily low while maintaining sufficient filter performance, as the suppression of carrier light at a frequency $\nu$ (corresponding to $\epsilon = 0$) is reduced for $\epsilon_r \to 0$. Specifically, below an offset frequency of $\epsilon_r^{\min} \approx 8$ MHz, the filters no longer have sufficiently rapid roll-off to suppress carrier photons, giving a rate of carrier photons reaching the photodetector greater than the flux from classical noise [$\dot{N}_{\text{pass}}^c \approx 1.6 \times 10^{-2}$ Hz, Eq. (14)]. Likewise, $\epsilon_r$ cannot be chosen arbitrarily high, as too much light in higher-order spatial modes is expected to be passed by the cavities at higher frequencies $\epsilon$. The filter cavities are designed to have a round-trip Gouy phase accumulation slightly offset from $2\pi/3$, in either direction. Thus, if the offset frequency from the carrier, $\epsilon_r$, is more than $1/3$ of the FSR of the cavity, carrier light in higher-order spatial modes will leak through. This sets $\epsilon_r^{\max} = \text{FSR}/3 \approx 40$ MHz.

### 10. Reduction of SNSPD dark counts and blackbody background

The main challenge in the implementation of SNSPDs is the reduction of dark counts, i.e., signals in the absence of any light. The origin of intrinsic dark counts in SNSPDs is an ongoing topic of research but is expected to be thermally activated single-vortex crossing events [86,87], the rate of which has an exponential dependence on the bias current [88,89]. The intrinsic detection efficiency, however, saturates at a certain threshold bias current; it is, thus, advantageous to bias the detector at this point, where the dark count rate can be $< 10^{-5}$ Hz [37]. To reach sufficiently low intrinsic dark count rates, the SNSPD will be operated at a temperature $< 1$ K.

Care must also be taken in readout and bias electronics, which can increase the dark count rate in SNSPDs above the intrinsic levels. The use of cryogenic (4 K) amplifiers and bias tees can significantly reduce electronic noise coupling to the detector, as well as the use of a fully differential readout architecture [90], which minimizes electromagnetic interference and prevents ground loops in the circuitry.

It is important to ensure minimal coupling of background light to the photon detector as well as rejection of any blackbody radiation since SNSPDs optimized for 1550 nm can be sensitive to photon wavelengths $> 3.0$ μm. By using a single-mode optical fiber between the final, cryogenic, filter cavity and the SNSPD, the background light will be minimized. Ideally, the final filter cavity could also be located in a contiguous cryogenic radiation shield with the detector.

If additional short-pass and narrow-band filtering of the signal is deemed necessary, an effective approach could be implemented that has recently been demonstrated using custom free-space filters [41]. In this approach, light from the optical fiber would be collimated with a cryogenic lens and sent through a series of filters, prior to going to being focused onto the SNSPD with another cryogenic lens.

### 11. Additional subdominant noise sources

In addition to those considered above, we have considered the following noise sources: (environmental) seismic noise, noise from residual gas in the interferometer vacuum system, and quantum radiation pressure noise. While these are limiting noise sources for lower-frequency interferometers, the PSDs of many of these noises decrease with frequency and are, thus, very subdominant to mechanical thermal noise at 17.6 MHz. Radiation pressure noise can be significant at high frequencies, as it is proportional to the mechanical susceptibility of the mirrors, and this susceptibility has peaks at the frequencies of the longitudinal solid normal mode resonances. However, quantum radiation pressure is weaker than thermal-mechanical noise in the optics (see Appendix A 4 a) in our measurement band.

Another potentially relevant noise is due to air diffusion in the readout part of GQuEST (RAD), i.e., gas noise in the output beams of the interferometers that enters the filter cavities. The coupling mechanism of this noise is similar to that of the residual gas noise in the interferometer vacuum system. However, the important differences are that the air in the readout is at atmospheric pressure, and, therefore, the motion of air molecules is diffusive and the number density is drastically higher. Using the two-point correlator of molecule positions and the Green's function for the Fokker-Planck diffusion equation (Fick's second law), we derive the spectral density. We elide the derivation for conciseness, will include it in future work, and note that this expression agrees with prior numerical results [91] and analogous methodology for vapor cells [92–94]. We find the following noise PSD in the high-frequency limit:

$$S_L^{\text{RAD}}(\Omega) = \frac{4\pi\rho_{\text{air}}L_{\text{air}}D_{\text{air}}}{\Omega^2}\left(\frac{ka_{\text{air}}}{\pi\epsilon_0 w_{\text{air}}^2}\right)^2 \frac{P_{\text{out}}}{h\nu G}, \quad (A28)$$

where the number density of air $\rho_{\text{air}} = 2.7 \times 10^{25}$ m$^{-3}$, the mass diffusivity $D_{\text{air}} = 2 \times 10^{-5}$ m$^2$/s, the beam radius $w_{\text{air}} \approx 500$ μm, and $a_{\text{air}}/(4\pi\epsilon_0) = (n_{\text{air}}^2 - 1)/(4\pi\rho_{\text{air}}) \approx 2 \times 10^{-30}$ m$^3$ is the polarizability of air at the laser wavelength $\lambda$ according to the Lorentz-Lorenz relation. $L_{\text{air}} \approx 4$ m is the approximate path length of the output light through the air between an IFO beam splitter (BS-A/BS-B) and the output of the first readout filter cavity. The path length inside the first readout cavity is included, as it resonantly enhances air noise sidebands at the readout frequency as much as it attenuates the carrier light. In effect, extra air diffusion noise is produced equivalent to the noise produced by light propagating for the length of the cavity. After the first cavity, the carrier light is suppressed significantly, and noise imparted in the other cavities is negligible. The air diffusion noise $S_L^{\text{RAD}}$ is subdominant, as shown in Fig. 5, but has an unfortunate scaling of $w_{\text{air}}^{-4}$ for the beam size, influencing the readout design.

### 12. Coherent signal detection challenges

The scheme for coherent signal detection with two collocated interferometers, using photon counting, shown in Fig. 4 is novel and, therefore, entails challenges not encountered in previous experiments. As discussed in Sec. IV H, the scheme combines outputs from the two interferometers on a beam splitter (BS-C) and provides two new output channels: a null channel and a signal channel. The optical path length between the output ports of the two IFOs and BS-C can be varied and is chosen such that a signal of a certain frequency that is coherent between the two IFOs interferes constructively in one of the output ports of BS-C. Any fields that have no coherence between the two output IFO ports will be split equally into the output ports of BS-C, and an estimate of the coherent signal power can, therefore, be obtained by taking the difference between the photon fluxes measured in the signal and null channels.

However, if the outputs of the IFOs contain correlated noise sideband fields, these fields would coherently interfere at BS-C, either constructively or destructively, and would, therefore, lead to an over- or underestimate of the coherent signal power, respectively. In other words, correlated noise (CN) effectively manifests as a spurious positive or negative signal power in the coherent signal detection scheme. Additionally, we separately consider photons that leak through the readout cavities in an optical mode different from that of the signal, i.e., in higher-order transverse modes or through scattering processes. If the leakage photon flux is different for the signal (addition) and null (subtraction) channels, then that gives rise to a bias, which like CN manifests as spurious positive or negative signal power.

We can write the photon flux that would be measured by a single-photon detector in each channel as [see Ref. [28], Eqs. (126) and (127)]

$$\dot{N}_{\text{signal}} = 2C^\phi \dot{N}^\phi_{\text{pass}} + 2\dot{N}^{\text{CN}}_{\text{add}} + \dot{N}^{\text{leak}}_{\text{add}} \\ + \left[(1 - C^\phi)\dot{N}^\phi_{\text{pass}} + \dot{N}^{\text{UN}}\right], \quad \text{(A29)}$$

$$\dot{N}_{\text{null}} = 2\dot{N}^{\text{CN}}_{\text{sub}} + \dot{N}^{\text{leak}}_{\text{sub}} + \left[(1 - C^\phi)\dot{N}^\phi_{\text{pass}} + \dot{N}^{\text{UN}}\right], \quad \text{(A30)}$$

where $\dot{N}^{\text{UN}}$ is the photon flux due to uncorrelated noise passed by the filter cavities; note that both channels contain half of the uncorrelated noise and half of the incoherent part of the signal. $\dot{N}^{\text{CN}}_{\text{add,sub}}$ are the photon fluxes passed by the filter cavities due to the addition and subtraction of noise that is coherent between the interferometers (or between the input ports of BS-C). While the dominant thermal noises in the GQuEST experiment are expected to have no such coherence between the two IFOs, correlated noise might be inadvertently introduced by combining the IFO output fields, phase locking the respective input lasers, and radio-frequency electrical crosstalk. $\dot{N}^{\text{leak}}_{\text{add,sub}}$ are the photon fluxes of light in an optical mode different from the signal leaking through the filter cavities.

In the presence of CN and leakage photons, the bias in the measurement of the signal photon flux can be given as

$$\dot{N}^{\text{bias}} = 2\left(\dot{N}^{\text{CN}}_{\text{add}} - \dot{N}^{\text{CN}}_{\text{sub}}\right) + \left(\dot{N}^{\text{leak}}_{\text{add}} - \dot{N}^{\text{leak}}_{\text{sub}}\right). \quad \text{(A31)}$$

This bias photon flux must be estimated and subtracted to correct the measured signal photon flux and evaluate the SNR for the signal of interest in the coherent signal detection scheme:

$$\text{SNR}^2_{\text{biased coh cts}} = \int_0^T \frac{\left[\left(\dot{N}_{\text{signal}} - \dot{N}_{\text{null}} - \dot{N}^{\text{bias}}\right)dt\right]^2}{\left(\dot{N}_{\text{signal}} + \dot{N}_{\text{null}}\right)dt}. \quad \text{(A32)}$$

This expression, when expanded, is consistent with Eq. (20).

The accurate evaluation of the SNR requires the accurate subtraction of the bias $\dot{N}^{\text{bias}}$; in practice, this subtraction of the bias term will add non-negligible uncertainty to the measurement of the signal power. The coherent signal flux $C^\phi \dot{N}^\phi_{\text{pass}}$ can be resolved only when $2C^\phi \dot{N}^\phi_{\text{pass}} > \Delta \dot{N}^{\text{bias}}$, where $\Delta \dot{N}^{\text{bias}}$ is the uncertainty in the estimation of $\dot{N}^{\text{bias}}$. This additional uncertainty can be formally accounted for by composing the SNR as the sum of the variances due to the statistical error and the bias (where the variance is proportional to the inverse of the $\text{SNR}^2$). To this end, we first define $\text{SNR}^2_{\text{bias}}$ as the quadrature contribution to $\text{SNR}^2_{\text{coh counts}}$ from the bias photon flux:

$$\text{SNR}^2_{\text{bias}} = \int_0^T \frac{\left(2C^\phi \dot{N}^\phi_{\text{pass}}\right)^2}{\left(\Delta \dot{N}_{\text{bias}}\right)^2} dt. \quad \text{(A33)}$$

Thus, we have

$$\text{SNR}^{-2}_{\text{coh counts}} = \text{SNR}^{-2}_{\text{biased coh cts}} + \text{SNR}^{-2}_{\text{bias}}. \quad \text{(A34)}$$

We recall using Eq. (13) that $C^\phi \dot{N}^\phi_{\text{pass}} \approx C^\phi \alpha \cdot 1.4 \times 10^{-3}$ Hz. Thus, CN and photon leakage may undermine the coherent detection technique unless this correlated noise can be sufficiently mitigated or quantified precisely. We argue below that this is feasible for several potential sources of CN and photon leakage in nonsignal modes.

#### a. Sources of correlated noise

Sources of correlated noise may be categorized by whether they impart CN either upstream of BS-C, toward the IFO, or downstream, toward the output filter cavities; we first consider the former.

Input laser noise of the two interferometers is a potential source of CN. To ensure that the signal of interest (which is expected to be coherent across the two interferometers) creates coherent sideband signals, the carrier fields in the two interferometers need to be coherent as well. This necessitates phase locking the two input lasers together, which will be performed using a feedback control loop with a bandwidth of roughly 100 kHz. This has the drawback that noise introduced by this phase-lock feedback control system (controls noise) used to lock both lasers would be coherent between the instruments to a significant degree. However, since the control system will be operating at frequencies well below the nominal filter offset frequency $\epsilon_r$, the magnitude of correlated laser noise introduced thus is greatly reduced above the controller's bandwidth. In addition, the laser filter cavities and interferometer power-recycling cavities will further suppress this controls noise.

We now consider CN imparted at BS-C or any process downstream of BS-C. The power of noise sidebands produced there is proportional to the power of the incident light, and, therefore, the noise power produced at BS-C will be a factor of $P_{\text{out}}/P_{\text{BS}} = \mathcal{O}(10^{-5})$ lower compared to the noise produced at BS-A and BS-B. Because of this known scaling, many potential sources of CN in this part of the experiment can be characterized when operating a single interferometer. Moreover, by changing the relative phase between the two inputs of BS-C ($\Delta\Phi$ as depicted in Fig. 4), the signal and null channels can be switched between the physical output ports of the beam splitter, which allows the same readout cavities and detector system to be used to measure both channels (at different times) with minimal changes to the system. This avoids potential systematic errors (which would manifest in the same way as CN) in the measurement of the photon flux in the signal and null channels. The dominant source of CN imparted downstream of BS-C that is anticipated for the fiducial design is readout air diffusion noise (RAD; see Appendix A 11). We estimate, using Eq. (A28), that the CN photon flux from this source downstream of BS-C will be $\dot{N}^{\text{RAD}} = \mathcal{O}(4 \times 10^{-5})$ Hz. Given the expected scaling of this noise with experimental parameters that can be precisely varied, we expect to be able to characterize and subtract this noise to well within an order of magnitude. Moreover, this noise is not fundamental and can be removed almost entirely by reducing the output optical path length in the air or putting the first readout cavity in a vacuum chamber.

#### b. Sources of non-signal-mode photon leakage

A potentially significant source of leakage photons that is in different optical mode than the signal is output fringe light at the carrier frequency ($\epsilon = 0$), which is deliberately made coherent between the two IFOs as explained above and, therefore, interferes coherently at BS-C and preferentially goes into one of the outputs of BS-C. This light is suppressed by $\mathcal{O}(240 \text{ dB})$ through the filter cavities (see Sec. IV B), but any remaining light manifests as noise. This photon flux is characterized by modulating the output light level through changing the relative phase $\Delta\Phi$ between the two outputs of the IFOs, as well as by changing the output light level of either IFO. This testing process modulates spurious photon flux from the leakage of carrier light while not modulating flux from high-frequency thermal noise processes, which allows this photon leakage source to be appropriately measured and subtracted from the measured coherent signal power in the data. If the filter cavities achieve their fiducial design performance, the photon flux of carrier light in the fundamental mode leaking through the filter cavities is expected to be $\dot{N}_{\text{pass}}^{\text{carrier}} = \mathcal{O}(10^{-6})$ Hz.

A related concern is carrier light leaking through the filter cavities in HOMs. This photon leakage source can be characterized by introducing an aperture into the output beam of a single interferometer to scatter a significant amount of light into HOMs, which then allows the leakage of HOMs through the filter cavities to be measured. For the fiducial cavity design, where each cavity's FSR is slightly different, we expect the photon flux from carrier HOM leakage to be subdominant to fundamental mode carrier leakage at the fiducial readout frequency $\epsilon_r$. We leave precise modeling of the expected HOM leakage flux for future work.

## APPENDIX B: VALIDATION OF THE SNRS AND RELATION TO THE QUANTUM FISHER INFORMATION

Herein, we validate the expressions for the SNRs given above and relate them to the generalized quantum Fisher information analysis in Ref. [30]. We refer to Ref. [30] here as SQFI (for stochastic quantum Fisher information).

SQFI rigorously proves the optimality of photon counting for detecting stochastic signals encoded onto a vacuum quantum state by a quantum "channel." A channel is a formal abstraction for the action of an experiment on a quantum system. The proof proceeds by showing that photon-counting measurements saturate the available quantum Fisher information for the estimation of the parameter that defines a stochastic signal of interest; i.e., SQFI shows that photon-counting readout, as proposed for GQuEST, saturates the quantum Cramér-Rao bound for the signal process involved and is, therefore, an optimal measurement scheme. SQFI then proceeds to derive how to increase the Fisher information using optimal state preparation and determines that non-Gaussian states outperform squeezed states for stochastic signal estimation. Our work, by contrast, uses physical arguments to compute a statistical SNR using Eq. (6). We show that practical implementations of photon counting realistically outperform homodyne readout, but we have not established the universal optimality of the technique. We also did not derive results for quantum state preparation beyond the default vacuum states.

Below, we validate our SNR calculations using the proofs provided by SQFI. Moreover, we present relationships to include quantum squeezing in SNR calculations and indicate the potential for quantum state preparation for GQuEST as quoted in Sec. VI B. SQFI uses different conventions than we do, as that paper has a different goal and target audience in quantum information science. We start by relating the different conventions used in our paper to those used in SQFI. In our work, we seek to estimate the parameter $\alpha$, which parametrizes the magnitude of the signal power spectral density $S_L^\phi(f)$. SQFI alternately considers estimation of the standard deviation $\sigma(f)$ of a random signal in a single quantum state measurement. This implies $\alpha \propto S_L^\phi(f) \propto \sigma^2(f)$. For a broadband time-invariant signal, there is an associated signal variance $\sigma^2(f)$ at every measurement frequency. In SQFI, the parameter $\sigma(f)$ is normalized such that the variance in phase quadrature of a vacuum state, acted on by the experiment (i.e., the channel), is equal to $\sigma^2(f) + 1/2$. The constant $1/2$ term is the contribution from vacuum fluctuations. Thus, we can relate the signal power spectral density to the generalized frequency-dependent signal variance as

$$\sigma^2(f) \equiv \frac{S_L^\phi(f)}{2\bar{S}_L^q} = 2\mathcal{S}_N^\phi(\pm f). \tag{B1}$$

We further want to relate the Fisher information about the signal parameter $\mathcal{I}[\alpha]$ used in our work to the more generalized Fisher information $\mathcal{I}[\sigma(f)]$ about a random signal's standard deviation $\sigma(f)$ as considered in SQFI, so that we may apply the findings from SQFI to our SNR calculations. Note that the Fisher information considered in SQFI is for an estimate of the signal parameter at a given frequency $\sigma(f)$ or, equivalently, for a pair of states (see Sec. I in Ref. [30]). In contrast, our paper considers Fisher information integrated over a time-bandwidth product. This integration is implicit in performing a realistic measurement and is, therefore, included in our definition of the total Fisher information $\mathcal{I}_M[\alpha]$. Thus, our convention for the SNR also includes this accumulation over measurement time and frequency in Eq. (6). We can relate the generalized Fisher information $\mathcal{I}_M[\sigma^2(f)]$ considered in SQFI to $\mathcal{I}_M[\alpha]$ as used in our paper by integrating over frequency (or over states) within a time-bandwidth product and changing parameters from $\sigma(f)$ to $\alpha$:

$$\mathcal{I}_M[\alpha] \equiv 2\int_0^T dt \int_0^{\Delta f} df \left(\frac{d\sigma^2(f)}{d\alpha}\right)^2 \mathcal{I}_M[\sigma^2(f)]. \tag{B2}$$

The factor of 2 accounts for measuring (extracting information from) two quantum states at each offset frequency $f$. In a photon-counting context, this pair of measurements is related to the existence of two quantum states corresponding to the upper and lower $\epsilon = \pm f$ optical frequencies. In a homodyne context, the recorded time-series measures the two states through each of the independent $\cos(f)$ and $\sin(f)$ terms at every frequency [95].

To evaluate the prior expression, we compute the derivative of the frequency-dependent variance $\sigma^2(f)$ to the frequency-independent parameter $\alpha$ using Eq. (B1):

$$d\sigma^2(f) = \frac{1}{4\bar{S}_L^q}\frac{dS_L^\phi(f)}{d\alpha}d\alpha = \frac{S_L^\phi(f)}{4\bar{S}_L^q\alpha}d\alpha = \frac{\sigma^2(f)}{\alpha}d\alpha. \tag{B3}$$

We also use the standard reparametrization of the Fisher information into variance as given by SQFI in Sec. I B therein:

$$\mathcal{I}_M[\sigma(f)] = 4\sigma^2 \mathcal{I}_M[\sigma^2(f)]. \tag{B4}$$

The SNR as defined in Eq. (6) can now be directly related to the generalized Fisher information about the signal standard deviation $\mathcal{I}_M[\sigma(f)]$:

$$\text{SNR}_M^2 \equiv \alpha^2 \mathcal{I}_M[\alpha] \equiv 2T \int_0^{\Delta f} df \frac{\sigma^2(f)}{4} \mathcal{I}_M[\sigma(f)]. \tag{B5}$$

Using the equation above, we can use the expressions for the generalized Fisher information $\mathcal{I}_M[\sigma(f)]$ as given in SQFI for different measurement schemes $M$ to derive expressions for the SNR that account for quantum state preparation and losses.

We first evaluate the SNR for a homodyne measurement that uses quantum state preparation, where we account for the losses $\Lambda_{\text{sig}}$ along the signal path [96]. As given by SQFI, we have for the generalized Fisher information

$$\mathcal{I}_{\text{homodyneSQZ}}[\sigma(f)] = \frac{8\sigma^2}{\Lambda_{\text{sig}}^2}, \quad \text{assuming } \sigma^2(f) \ll \frac{1}{2}, \tag{B6}$$

which gives for the SNR in a homodyne readout scheme that may include squeezing and losses:

$$\text{SNR}_{\text{homodyneSQZ}}^2 = 2T \int_0^{\Delta f} 2\sigma^4(f) df \tag{B7}$$

$$= T \int_0^{\Delta f} \left(\frac{S_L^\phi(f)}{\Lambda_{\text{sig}}\bar{S}_L^q}\right)^2 df \tag{B8}$$

$$\approx T\Delta f \frac{1}{(\Lambda_{\text{sig}})^2}\left(\frac{\bar{S}_L^\phi}{\bar{S}_L^q}\right)^2. \tag{B9}$$

Setting $\Lambda_{\text{sig}} = 1$ corresponds to 100% loss of the injected squeezed state and, thus, recovers the SNR for homodyne readout using the default vacuum state injection [see Eq. (7)]. This result was originally computed by Price and Middleton in the context of classical Gaussian noise [31,32]. Setting $\Lambda_{\text{sig}} < 1$ recovers the enhancement quoted in Sec. VI B from squeezed state injection.

We continue by analyzing photon-counting readout. This requires including the limitations arising from classical noise. This noise source is both practically and formally significant for photon-counting readout, in contrast to shot-noise-limited homodyne readout. Its practical significance is for the reasons developed in Sec. IV G. Its formal significance is that not accounting for the classical noise causes the quantum Fisher information about the variance to diverge for small signals. Sections I B and II of SQFI discuss the significance of this divergence and the related concepts of quantum super-resolution and the "Rayleigh curse."

Background noise is parametrized in SQFI as $\sigma_p(f)$, which is defined as the standard deviation of noise in the phase quadrature of the output state due to classical processes. We can relate $\sigma_p(f)$ to the classical noise power spectral density $S_L^c(f)$ similarly to Eq. (B1), which gives $\sigma_p^2(f) = S_L^c(f)/(2\bar{S}_L^q)$.

The Fisher information for photon-counting readout $\mathcal{I}_{\text{counts}}[\sigma]$ is provided by Eq. (27) of SQFI. This is computed using the extended channel quantum Fisher information (ECQFI), which determines the maximum available information extraction over all possible quantum state preparations and ancilla channels over which states may be prepared and measured [97].

In the limit where the signal and the classical noise are small (i.e., the loss-dominated limit where $0 < \sigma^2(f) \ll \sigma_p^2(f) \ll \Lambda_{\text{sig}}$), the ECQFI and the Fisher information for photon counting are given, respectively, by

$$\mathcal{I}_{\text{ECQFI}}[\sigma(f)] = \frac{2\sigma^2(f)}{\Lambda_{\text{sig}}\sigma_p^2(f)}, \quad \mathcal{I}_{\text{counts}}[\sigma] = \frac{2\sigma^2(f)}{\sigma_p^2(f)}. \quad (B10)$$

This shows that $\mathcal{I}_{\text{counts}}$ saturates the ECQFI when default vacuum states are injected, as this is equivalent to $\Lambda_{\text{sig}} = 1$. When losses are small, $\mathcal{I}_{\text{counts}}$ does not saturate the ECQFI, and a combination of state preparation and more general quantum measurement process is optimal (see SQFI Sec. VI B). We emphasize that such a more general measurement process is likely to be practically implemented more similarly to photon counting than to homodyne readout, utilizing and benefiting from the experimental analysis of this work.

Using Eq. (B10) in Eq. (B5) allows us to evaluate the optimal SNR according to the ECQFI, including the effect of quantum state preparation limited by losses:

$$\text{SNR}_{\text{ECQFI}}^2 = 2T \int_0^{\Delta f} \frac{\sigma^4(f)}{2\Lambda_{\text{sig}}\sigma_p^2(f)} df \quad (B11)$$

$$= \frac{T}{\Lambda_{\text{sig}}} \int_0^{\Delta f} \frac{(S_L^\phi(f))^2}{2S_L^c(f)\bar{S}_L^q} df \quad (B12)$$

$$= \frac{T}{\Lambda_{\text{sig}}} \int_{-\Delta f}^{\Delta f} \frac{(S_L^\phi(\epsilon))^2}{4S_L^c(\epsilon)\bar{S}_L^q} d\epsilon. \quad (B13)$$

We can now make a direct quantitative comparison of the SNR of homodyne vs a generalized photon-counting readout that saturates the ECQFI, including quantum state preparation for both schemes. We, therefore, evaluate the ratio of the integrands of the two SNRs, assuming integration of the same frequency band and including quantum information from both upper and lower sideband frequencies:

$$\frac{\text{SNR}_{\text{ECQFI}}^2}{\text{SNR}_{\text{homodyneSQZ}}^2} \approx \Lambda_{\text{sig}} \frac{\bar{S}_L^q}{2\bar{S}_L^c}. \quad (B14)$$

For GQuEST, the factor $\bar{S}_L^q/(2\bar{S}_L^c) = \mathcal{O}(5 \times 10^6)$, which indicates that homodyne readout, even when combined with realistic levels of squeezing, is significantly less effective than the ECQFI indicates is possible for more general state preparation and measurement.

We can further compare the SNR for photon-counting readout of the fiducial GQuEST design, i.e., including filter cavities without quantum state preparation, to homodyne readout with squeezing. For that comparison, we adjust the SNR ratio above to account for the reduction in measured signal bandwidth and the lack of quantum state preparation:

$$\text{SNR}_{\text{counts}}^2 \approx T\Delta\epsilon \frac{(\bar{S}_L^\phi)^2}{4\bar{S}_L^c \bar{S}_L^q}. \quad (B15)$$

This approximation is related to $\text{SNR}_{\text{ECQFI}}^2$ by setting $\Lambda_{\text{sig}} = 1$ and restricting the limits of integration to a band of width $\Delta\epsilon$; we, thus, recover Eq. (19). For smaller values of $\Lambda_{\text{sig}}$, corresponding to the injection of prepared quantum states, the SNR can be improved to the ECQFI limit. The results for the improvements that can be obtained are quoted in Sec. VI B.

The relative difference of the SNR of photon counting as in the fiducial design and homodyne readout with squeezing, around $f_{\text{peak}}$, is

$$\frac{\text{SNR}_{\text{counts}}^2}{\text{SNR}_{\text{homodyneSQZ}}^2} \approx \Lambda_{\text{sig}}^2 \frac{\bar{S}_L^q}{2\bar{S}_L^c}\left(\frac{\Delta\epsilon}{2\Delta f}\right). \quad (B16)$$

This shows the relatively greater improvement from squeezing with respect to vacuum states gained by homodyne readout. For GQuEST, the additional bandwidth factor in parentheses causes the SNR ratio to amount to roughly a factor of $70\Lambda_{\text{sig}}^2$ of measurement time reduction for a photon counting vs a homodyne readout design; i.e., the fiducial photon-counting readout design breaks even with homodyne readout and squeezing at $\Lambda_{\text{sig}} \approx 12\%$. If future technology enables photon counting over a greater bandwidth $\Delta\epsilon$ and using more general quantum state preparation, the advantage of photon counting over homodyne readout can increase and approach the ratio given by Eq. (B14).

---

[1] Y. Jack Ng and H. Van Dam, *Limit to space-time measurement*, Mod. Phys. Lett. A **09**, 335 (1994).

[2] G. Amelino-Camelia, *Limits on the measurability of space-time distances in (the semi-classical approximation of) quantum gravity*, Mod. Phys. Lett. A **09**, 3415 (1994).

[3] C. J. Hogan, *Measurement of quantum fluctuations in geometry*, Phys. Rev. D **77**, 104031 (2008).

[4] E. P. Verlinde and K. M. Zurek, *Observational signatures of quantum gravity in interferometers*, Phys. Lett. B **822**, 136663 (2021).

[5] O. Kwon, *Phenomenology of holography via quantum coherence on causal horizons*, arXiv:2204.12080.

[6] K. M. Zurek, *Snowmass 2021 white paper: Observational signatures of quantum gravity*, arXiv:2205.01799.

[7] K. M. Zurek, *On vacuum fluctuations in quantum gravity and interferometer arm fluctuations*, Phys. Lett. B **826**, 136910 (2022).

[8] E. Verlinde and K. M. Zurek, *Modular fluctuations from shockwave geometries*, Phys. Rev. D **106**, 106011 (2022).

[9] T. He, A.-M. Raclariu, and K. M. Zurek, *From shockwaves to the gravitational memory effect*, J. High Energy Phys. 01 (2024) 006.

[10] D. Li, V. S. H. Lee, Y. Chen, and K. M. Zurek, *Interferometer response to geontropic fluctuations*, Phys. Rev. D **107**, 024002 (2023).

[11] M. W. Bub, Y. Chen, Y. Du, D. Li, Y. Zhang, and K. M. Zurek, *Quantum gravity background in next-generation gravitational wave detectors*, Phys. Rev. D **108**, 064038 (2023).

[12] T. Jacobson, *Entanglement equilibrium and the Einstein equation*, Phys. Rev. Lett. **116**, 201101 (2016).

[13] H. Casini, M. Huerta, and R. C. Myers, *Towards a derivation of holographic entanglement entropy*, J. High Energy Phys. 05 (2011) 036.

[14] T. Banks and K. M. Zurek, *Conformal description of near-horizon vacuum states*, Phys. Rev. D **104**, 126026 (2021).

[15] S. Gukov, V. S. H. Lee, and K. M. Zurek, *Near-horizon quantum dynamics of 4D Einstein gravity from 2D Jackiw-Teitelboim gravity*, Phys. Rev. D **107**, 016004 (2023).

[16] Y. Zhang and K. M. Zurek, *Stochastic description of near-horizon fluctuations in Rindler-AdS*, Phys. Rev. D **108**, 066002 (2023).

[17] E. Verlinde and K. M. Zurek, *Spacetime fluctuations in AdS/CFT*, J. High Energy Phys. 04 (2020) 209.

[18] A. Buikema, C. Cahillane, G. L. Mansell et al., *Sensitivity and performance of the Advanced LIGO detectors in the third observing run*, Phys. Rev. D **102**, 062003 (2020).

[19] A. Chou, H. Glass, H. R. Gustafson, C. Hogan, B. L. Kamai, O. Kwon, R. Lanza, L. McCuller, S. S. Meyer, J. Richardson, C. Stoughton, R. Tomlin, and R. Weiss, *The holometer: An instrument to probe Planckian quantum geometry*, Classical Quantum Gravity **34**, 065005 (2017).

[20] S. M. Vermeulen, L. Aiello, A. Ejlli, W. L. Griffiths, A. L. James, K. L. Dooley, and H. Grote, *An experiment for observing quantum gravity phenomena using twin table-top 3d interferometers*, Classical Quantum Gravity **38**, 085008 (2021).

[21] W. A. Christiansen, Y. J. Ng, D. J. E. Floyd, and E. S. Perlman, *Limits on spacetime foam*, Phys. Rev. D **83**, 084003 (2011).

[22] E. S. Perlman, S. A. Rappaport, W. A. Christensen, Y. J. Ng, J. DeVore, and D. Pooley, *New constraints on quantum gravity from x-ray and gamma-ray observations*, Astrophys. J. **805**, 10 (2015).

[23] Y. J. Ng and E. S. Perlman, *Probing spacetime foam with extragalactic sources of high-energy photons*, Universe **8**, 382 (2022).

[24] V. S. H. Lee, K. M. Zurek, and Y. Chen, *Astronomical image blurring from transversely correlated quantum gravity fluctuations*, Phys. Rev. D **109**, 084005 (2024).

[25] P. Fritschel, M. Evans, and V. Frolov, *Balanced homodyne readout for quantum limited gravitational wave detectors*, Opt. Express **22**, 4224 (2014).

[26] C. M. Caves, *Quantum-mechanical noise in an interferometer*, Phys. Rev. D **23**, 1693 (1981).

[27] A. Buonanno, Y. Chen, and N. Mavalvala, *Quantum noise in laser-interferometer gravitational-wave detectors with a heterodyne readout scheme*, Phys. Rev. D **67**, 122005 (2003).

[28] L. McCuller, *Single-photon signal sideband detection for high-power Michelson interferometers*, arXiv:2211.04016.

[29] Note that the SNR in Eq. (7) turns out to be the ratio of signal power to noise power for the case of homodyne readout, but this is not true by definition. Our definition is purely statistical; cf. Eq. (15), where the SNR for photon-counting readout is found to be the ratio of the signal photon flux to the square root of the total photon flux.

[30] J. W. Gardner, T. Gefen, S. A. Haine, J. J. Hope, J. Preskill, Y. Chen, and L. McCuller, *Stochastic waveform estimation at the fundamental quantum limit*, arXiv:2404.13867.

[31] R. Price, *Optimum detection of random signals in noise, with application to scatter-multipath communication—I*, IRE Trans. Inf. Theory **2**, 125 (1956).

[32] D. Middleton, *On the detection of stochastic signals in additive normal noise—I*, IRE Trans. Inf. Theory **3**, 86 (1957).

[33] E. E. Flanagan, *Sensitivity of the Laser Interferometer Gravitational Wave Observatory to a stochastic background, and its dependence on the detector orientations*, Phys. Rev. D **48**, 2389 (1993).

[34] S. Ng, S. Z. Ang, T. A. Wheatley, H. Yonezawa, A. Furusawa, E. H. Huntington, and M. Tsang, *Spectrum analysis with quantum dynamical systems*, Phys. Rev. A **93**, 042121 (2016).

[35] F. Meylahn and B. Willke, *Characterization of laser systems at 1550 nm wavelength for future gravitational wave detectors*, Instruments **6**, 15 (2022).

[36] D. V. Reddy, R. R. Nerem, S. W. Nam, R. P. Mirin, and V. B. Verma, *Superconducting nanowire single-photon detectors with 98% system detection efficiency at 1550 nm*, Optica **7**, 1649 (2020).

[37] J. Chiles, I. Charaev, R. Lasenby, M. Baryakhtar, J. Huang, A. Roshko, G. Burton, M. Colangelo, K. Van Tilburg, A. Arvanitaki, S. W. Nam, and K. K. Berggren, *New constraints on dark photon dark matter with superconducting nanowire detectors in an optical haloscope*, Phys. Rev. Lett. **128**, 231802 (2022).

[38] F. Marsili, V. B. Verma, J. A. Stern, S. Harrington, A. E. Lita, T. Gerrits, I. Vayshenker, B. Baek, M. D. Shaw, R. P. Mirin, and S. W. Nam, *Detecting single infrared photons with 93% system efficiency*, Nat. Photonics **7**, 210 (2013).

[39] D. Y. Vodolazov, *Single-photon detection by a dirty current-carrying superconducting strip based on the kinetic-equation approach*, Phys. Rev. Appl. **7**, 034014 (2017).

[40] J. P. Allmaras, A. G. Kozorezov, B. A. Korzh, K. K. Berggren, and M. D. Shaw, *Intrinsic timing jitter and latency in superconducting nanowire single-photon detectors*, Phys. Rev. Appl. **11**, 034062 (2019).

[41] A. S. Mueller, B. Korzh, M. Runyan, E. E. Wollman, A. D. Beyer, J. P. Allmaras, A. E. Velasco, I. Craiciu, B. Bumble, R. M. Briggs, L. Narvaez, C. Peña, M. Spiropulu, and M. D. Shaw, *Free-space coupled superconducting nanowire single-photon detector with low dark counts*, Optica **8**, 1586 (2021).

[42] R. X. Adhikari *et al.*, *A cryogenic silicon interferometer for gravitational-wave detection*, Classical Quantum Gravity **37**, 165003 (2020).

[43] T. Hong, H. Yang, E. K. Gustafson, R. X. Adhikari, and Y. Chen, *Brownian thermal noise in multilayer coated mirrors*, Phys. Rev. D **87**, 082001 (2013).

[44] B. Benthem and Y. Levin, *Thermorefractive and thermochemical noise in the beam splitter of the GEO600 gravitational-wave interferometer*, Phys. Rev. D **80**, 062004 (2009).

[45] S. M. Vermeulen, *Fundamental physics with laser interferometry*, Ph.D. thesis, Cardiff University, 2023.

[46] A. Chou, H. Glass, H. R. Gustafson, C. J. Hogan, B. L. Kamai, O. Kwon, R. Lanza, L. McCuller, S. S. Meyer, J. W. Richardson, C. Stoughton, R. Tomlin, and R. Weiss, *Interferometric constraints on quantum geometrical shear noise correlations*, Classical Quantum Gravity **34**, 165005 (2017).

[47] B. J. Meers, *Recycling in laser-interferometric gravitational-wave detectors*, Phys. Rev. D **38**, 2317 (1988).

[48] S. Hild, H. Grote, M. Hewitson, H. Lück, J. R. Smith, K. A. Strain, B. Willke, and K. Danzmann, *Demonstration and comparison of tuned and detuned signal recycling in a large-scale gravitational wave detector*, Classical Quantum Gravity **24**, 1513 (2007).

[49] D. E. McClelland, *An overview of recycling in laser interferometric gravitational wave detectors.*, Aust. J. Phys. **48**, 953 (1995).

[50] S. Hild, H. Grote, J. Degallaix, S. Chelkowski, K. Danzmann, A. Freise, M. Hewitson, J. Hough, H. Lueck, M. Prijatelj, K. A. Strain, J. R. Smith, and B. Willke, *DC-readout of a signal-recycled gravitational wave detector*, Classical Quantum Gravity **26**, 055012 (2009).

[51] D. Martynov, E. Hall, B. Abbott *et al.*, *Sensitivity of the advanced LIGO detectors at the beginning of gravitational wave astronomy*, Phys. Rev. D **93**, 112004 (2016).

[52] S. M. Vermeulen, L. Aiello, A. Ejlli, W. L. Griffiths, A. L. James, K. L. Dooley, and H. Grote, *An experiment for observing quantum gravity phenomena using twin table-top 3D interferometers*, Classical Quantum Gravity **38**, 085008 (2021).

[53] C. L. Latune, B. M. Escher, R. L. de Matos Filho, and L. Davidovich, *Quantum limit for the measurement of a classical force coupled to a noisy quantum-mechanical oscillator*, Phys. Rev. A **88**, 042112 (2013).

[54] R. Demkowicz-Dobrzański, K. Banaszek, and R. Schnabel, *Fundamental quantum interferometry bound for the squeezed-light-enhanced gravitational wave detector GEO 600*, Phys. Rev. A **88**, 041802(R) (2013).

[55] Note that Eq. (7) is for a shot-noise-dominated measurement, where any classical noise is neglected. Classical noise would ultimately limit a homodyne measurement that uses high levels of squeezing.

[56] J. Lough, E. Schreiber, F. Bergamin *et al.*, *First demonstration of 6 dB quantum noise reduction in a kilometer scale gravitational wave observatory*, Phys. Rev. Lett. **126**, 041102 (2021).

[57] D. Ganapathy, W. Jia, M. Nakano *et al.*, *Broadband quantum enhancement of the LIGO detectors with frequency-dependent squeezing*, Phys. Rev. X **13**, 041021 (2023).

[58] M. Maggiore, C. V. D. Broeck, N. Bartolo *et al.*, *Science case for the Einstein telescope*, J. Cosmol. Astropart. Phys. 03 (2020) 050.

[59] M. Evans, R. X. Adhikari, C. Afle *et al.*, *A horizon study for Cosmic Explorer: Science, observatories, and community*, arXiv:2109.09882.

[60] S. M. Vermeulen, P. Relton, H. Grote *et al.*, *Direct limits for scalar field dark matter from a gravitational-wave detector*, Nature (London) **600**, 424 (2021).

[61] D. Antypas, A. Banerjee, C. Bartram *et al.*, *New horizons: Scalar and vector ultralight dark matter*, arXiv:2203.14915.

[62] R. Abbott, T. Abbott *et al.*, (Virgo Collaboration, LIGO Scientific Collaboration, and KAGRA Collaboration), *Constraints on dark photon dark matter using data from LIGO's and Virgo's third observing run*, Phys. Rev. D **105**, 063030 (2022).

[63] Y. Du, V. S. H. Lee, Y. Wang, and K. M. Zurek, *Macroscopic dark matter detection with gravitational wave experiments*, Phys. Rev. D **108**, 122003 (2023).

[64] C. Cahillane, G. L. Mansell, and D. Sigg, *Laser frequency noise in next generation gravitational-wave detectors*, Opt. Express **29**, 42144 (2021).

[65] P. Fritschel, M. Evans, and V. Frolov, *Balanced homodyne readout for quantum limited gravitational wave detectors*, Opt. Express **22**, 4224 (2014).

[66] P. R. Saulson, *Thermal noise in mechanical experiments*, Phys. Rev. D **42**, 2437 (1990).

[67] W. Yam, S. Gras, and M. Evans, *Multimaterial coatings with reduced thermal noise*, Phys. Rev. D **91**, 042002 (2015).

[68] H. B. Callen and T. A. Welton, *Irreversibility and generalized noise*, Phys. Rev. **83**, 34 (1951).

[69] Y. Levin, *Internal thermal noise in the LIGO test masses: A direct approach*, Phys. Rev. D **57**, 659 (1998).

[70] A. Gillespie and F. Raab, *Thermally excited vibrations of the mirrors of laser interferometer gravitational-wave detectors*, Phys. Rev. D **52**, 577 (1995).

[71] J. Rodriguez, S. A. Chandorkar, C. A. Watson, G. M. Glaze, C. H. Ahn, E. J. Ng, Y. Yang, and T. W. Kenny, *Direct detection of Akhiezer damping in a silicon MEMS resonator*, Sci. Rep. **9**, 2244 (2019).

[72] G. M. Harry, A. M. Gretarsson, P. R. Saulson, S. E. Kittelberger, S. D. Penn, W. J. Startin, S. Rowan, M. M. Fejer, D. R. M. Crooks, G. Cagnoli, J. Hough, and N. Nakagawa, *Thermal noise in interferometric gravitational wave detectors due to dielectric optical coatings*, Classical Quantum Gravity **19**, 897 (2002).

[73] S. Gras and M. Evans, *Direct measurement of coating thermal noise in optical resonators*, Phys. Rev. D **98**, 122001 (2018).

[74] V. B. Braginsky and S. P. Vyatchanin, *Corner reflectors and quantum-non-demolition measurements in gravitational wave antennae*, Phys. Lett. A **324**, 345 (2004).

[75] V. Braginsky, M. Gorodetsky, and S. Vyatchanin, *Thermodynamical fluctuations and photo-thermal shot noise in gravitational wave antennae*, Phys. Lett. A **264**, 1 (1999).

[76] M. Evans, S. Ballmer, M. Fejer, P. Fritschel, G. Harry, and G. Ogin, *Thermo-optic noise in coated mirrors for high-precision optical measurements*, Phys. Rev. D **78**, 102003 (2008).

[77] V. B. Braginsky, M. L. Gorodetsky, and S. P. Vyatchanin, *Thermo-refractive noise in gravitational wave antennae*, Phys. Lett. A **271**, 303 (2000).

[78] V. B. Braginsky, M. L. Gorodetsky, and S. P. Vyatchanin, *Thermodynamical fluctuations and photo-thermal shot noise in gravitational wave antennae*, Phys. Lett. A **264**, 1 (1999).

[79] M. M. Fejer, S. Rowan, G. Cagnoli, D. R. M. Crooks, A. Gretarsson, G. M. Harry, J. Hough, S. D. Penn, P. H. Sneddon, and S. P. Vyatchanin, *Thermoelastic dissipation in inhomogeneous media: Loss measurements and displacement noise in coated test masses for interferometric gravitational wave detectors*, Phys. Rev. D **70**, 082003 (2004).

[80] K. Somiya and K. Yamamoto, *Coating thermal noise of a finite-size cylindrical mirror*, Phys. Rev. D **79**, 102004 (2009).

[81] H. Siegel and Y. Levin, *Revisiting thermal charge carrier refractive noise in semiconductor optics for gravitational-wave interferometers*, Phys. Rev. D **107**, 022002 (2023).

[82] F. Bruns, S. P. Vyatchanin, J. Dickmann, R. Glaser, D. Heinert, R. Nawrodt, and S. Kroker, *Thermal charge carrier driven noise in transmissive semiconductor optics*, Phys. Rev. D **102**, 022006 (2020).

[83] K. A. Strain, K. Danzmann, J. Mizuno, P. G. Nelson, A. Rüdiger, R. Schilling, and W. Winkler, *Thermal lensing in recycling interferometric gravitational wave detectors*, Phys. Lett. A **194**, 124 (1994).

[84] M. R. Ardigo, M. Ahmed, and A. Besnard, *Stoney formula: Investigation of curvature measurements by optical profilometer*, Adv. Mater. Res. **996**, 361 (2014).

[85] L. McCuller, *LIGO-t1900144-v3: Beam layout requirements imposed by wavefront actuators*, LIGO Technical Note No. LIGO-T1900144-v3.

[86] L. N. Bulaevskii, M. J. Graf, C. D. Batista, and V. G. Kogan, *Vortex-induced dissipation in narrow current-biased thin-film superconducting strips*, Phys. Rev. B **83**, 144526 (2011).

[87] L. N. Bulaevskii, M. J. Graf, and V. G. Kogan, *Vortex-assisted photon counts and their magnetic field dependence in single-photon superconducting detectors*, Phys. Rev. B **85**, 014505 (2012).

[88] H. Bartolf, A. Engel, A. Schilling, K. Il'in, M. Siegel, H.-W. Hübers, and A. Semenov, *Current-assisted thermally activated flux liberation in ultrathin nanopatterned NbN superconducting meander structures*, Phys. Rev. B **81**, 024502 (2010).

[89] T. Yamashita, S. Miki, K. Makise, W. Qiu, H. Terai, M. Fujiwara, M. Sasaki, and Z. Wang, *Origin of intrinsic dark count in superconducting nanowire single-photon detectors*, Appl. Phys. Lett. **99**, 161105 (2011).

[90] M. Colangelo, B. Korzh, J. P. Allmaras et al., *Impedance-matched differential superconducting nanowire detectors*, Phys. Rev. Appl. **19**, 044093 (2023).

[91] R. Weiss, *LIGO-T2200336-v2: Considerations of a LIGO in air*, LIGO Technical Note No. LIGO-T2200336-v2.

[92] K. Aoki and T. Mitsui, *Observing random walks of atoms in buffer gas through resonant light absorption*, Phys. Rev. A **94**, 012703 (2016).

[93] V. G. Lucivero, N. D. McDonough, N. Dural, and M. V. Romalis, *Correlation function of spin noise due to atomic diffusion*, Phys. Rev. A **96**, 062702 (2017).

[94] S. Micalizio, A. Godone, M. Gozzelino, and F. Levi, *Brownian motion-induced amplitude noise in vapor-cell frequency standards*, New J. Phys. **22**, 083050 (2020).

[95] C. M. Caves and B. L. Schumaker, *New formalism for two-photon quantum optics. I. Quadrature phases and squeezed states*, Phys. Rev. A **31**, 3068 (1985).

[96] SQFI uses $\eta$ for this parameter; see Eq. (22) in Ref. [30].

[97] SQFI establishes that the ECQFI can be saturated without the use of ancilla channels by numerically computing effective states that approach the bound but do not use entanglement over an ancilla.